\documentclass[
	journal,
	letterpaper,
	twoside,
	twocolumn
]{IEEEtran}

\IEEEoverridecommandlockouts % This is to make the \thanks{} command visible in conference mode
\makeatletter
\def\markboth#1#2{\def\leftmark{\@IEEEcompsoconly{\sffamily}\MakeUppercase{\protect#1}}%
\def\rightmark{\@IEEEcompsoconly{\sffamily}\MakeUppercase{\protect#2}}}
\makeatother

\usepackage[amssymb]{SIunits}
\usepackage[cmex10]{amsmath} 
\usepackage{amssymb, amsthm}
\usepackage{bbm} % indicator function \mathbbm{1}
\usepackage[latin1]{inputenc}
\usepackage{xcolor}
\usepackage[noadjust]{cite} % noadjust: prevent inserting white spaces 
\usepackage[pdftex]{graphicx}
\usepackage[caption=false, font=footnotesize]{subfig}
\usepackage{url}
\usepackage{mathtools}
\DeclarePairedDelimiter{\ceil}{\lceil}{\rceil}

\usepackage{xspace} % for the ``space after newcommand'' problem
\usepackage{textcomp} % /textquotesingle mathmode

\usepackage{booktabs} % for \toprule, \midrule and \bottomrule macros

\usepackage{array}
\newcolumntype{L}[1]{>{\raggedright\let\newline\\\arraybackslash\hspace{0pt}}m{#1}}
\newcolumntype{C}[1]{>{\centering\let\newline\\\arraybackslash\hspace{0pt}}m{#1}}
\newcolumntype{R}[1]{>{\raggedleft\let\newline\\\arraybackslash\hspace{0pt}}m{#1}}

\usepackage[
	ruled,
	vlined,
	boxed, 
	linesnumbered, 
	commentsnumbered]{algorithm2e}

\usepackage[
	shortcuts,
	nonumberlist,	% don't show page numbers
	acronym, % use acronyms
	hyperfirst=true,
	section=section, % section-Ebene
	order=letter,
	numberedsection=nolabel % false / nolabel / autolabel
]{glossaries}

\theoremstyle{definition}

\theoremstyle{plain}

\theoremstyle{remark} % subsequently defined environments use this style
\newtheorem{remark}{Remark}
\newtheorem{example}{Example}

\newcommand\xqed[1]{%
\leavevmode\unskip\penalty9999 \hbox{}\nobreak\hfill
\quad\hbox{#1}}
\newcommand\demo{\xqed{$\triangle$}}
\newcommand{\IID}{i.i.d.\ }	% independent and indentically distributed	
\newcommand{\define}{\triangleq}

\newcommand{\dmin}{\ensuremath{d_\text{min}}}
\newcommand{\smin}{\ensuremath{s_\text{min}}}

\newcommand{\ErdRen}{Erd\H{o}s--R\'enyi\xspace}

\newcommand{\transpose}{\intercal}

\newcommand{\vect}[1]{\ensuremath{\boldsymbol{#1}}}
\newcommand{\mat}[1]{\ensuremath{\boldsymbol{#1}}}
\newcommand{\etab}{\ensuremath{\boldsymbol{\eta}}}

\newcommand{\GF}[1]{\ensuremath{\mathbb{F}_{#1}}}

\newcommand{\evec}{\ensuremath{\vect{n}}} 

\newif\ifshow
\showfalse

\newcommand{\abbr}[1]{{#1}}				

\makeatletter
\let\aclOLD=\acl
\renewcommand{\acl}[1]{%
  \begingroup    
  \let\@@underline=\relax
  \aclOLD{#1}%
  \endgroup
}
\makeatother

\newcommand{\NewA}[3]{
	\newacronym{#1}{#2}{#3}
}

\NewA{rbp}{RBP}{residual belief propagation}

\NewA{af}{AF}{\abbr{a}mplify-and-\abbr{f}orward}
\NewA{apsk}{APSK}{\abbr{a}mplitude \abbr{p}hase-\abbr{s}hift \abbr{k}eying}
\NewA{ask}{ASK}{\abbr{a}mplitude-\abbr{s}hift \abbr{k}eying}
\NewA{ase}{ASE}{\abbr{a}mplified \abbr{s}pontaneous \abbr{e}mission}
\NewA{awgn}{AWGN}{\abbr{a}dditive \abbr{w}hite \abbr{G}aussian \abbr{n}oise}
\NewA{biawgn}{BI-AWGN}{binary-input \abbr{a}dditive \abbr{w}hite \abbr{G}aussian \abbr{n}oise}
\NewA{bep}{BEP}{\abbr{b}it \abbr{e}rror \abbr{p}robability}
\NewA{cep}{CEP}{\abbr{c}odeword \abbr{e}rror \abbr{p}robability}
\NewA{ber}{BER}{\abbr{b}it \abbr{e}rror \abbr{r}ate}
\NewA{pp}{PP}{post-processing}
\NewA{qap}{QAP}{\abbr{q}uadratic \abbr{a}assignment \abbr{p}roblem}
\NewA{bicm}{BICM}{\abbr{b}it-\abbr{i}nterleaved \abbr{c}oded \abbr{m}odulation}				
\NewA{cm}{CM}{\abbr{c}oded \abbr{m}odulation}				
\NewA{qpsk}{QPSK}{quadrature phase-shift keying}				
\NewA{mlcm}{MLCM}{multilevel coded modulation}				
\NewA{bpsk}{BPSK}{\abbr{b}inary \abbr{p}hase-\abbr{s}hift \abbr{k}eying}				
\NewA{bsc}{BSC}{\abbr{b}inary \abbr{s}ymmetric \abbr{c}hannel}				
\NewA{brgc}{BRGC}{\abbr{b}inary \abbr{r}eflected \abbr{G}ray \abbr{c}ode}				
\NewA{cf}{CF}{\abbr{c}haracteristic \abbr{f}unction}
\NewA{csit}{CSIT}{\abbr{c}annnel \abbr{s}tate \abbr{i}nformation at the \abbr{transmitter}}
\NewA{csi}{CSI}{\abbr{c}annnel \abbr{s}tate \abbr{i}nformation}
\NewA{df}{DF}{\abbr{d}ecode-and-\abbr{f}orward}
\NewA{fd}{FD}{\abbr{f}ull-\abbr{d}uplex}
\NewA{fft}{FFT}{\abbr{f}ast \abbr{F}ourier \abbr{t}ransform}
\NewA{iid}{IID}{independent and \abbr{i}dentically \abbr{d}istributed}
\NewA{isi}{ISI}{\abbr{i}nter\abbr{s}ymbol \abbr{i}nterference}
\NewA{lb}{LB}{\abbr{l}ower \abbr{b}ound}
\NewA{map}{MAP}{\abbr{m}aximum \abbr{a} \abbr{p}osteriori}
\NewA{mf}{MF}{\abbr{m}odulo-and-\abbr{f}orward}
\NewA{mlan}{MLAN}{\abbr{m}odulo-\abbr{l}attice \abbr{a}dditive \abbr{n}oise}
\NewA{ml}{ML}{\abbr{m}aximum \abbr{l}ikelihood}
\NewA{mmse}{MMSE}{\abbr{m}inimum \abbr{m}ean \abbr{s}quare \abbr{e}rror}
\NewA{nlpc}{NLPC}{\abbr{n}onlinear \abbr{p}hase \abbr{c}ompensation}
\NewA{nlpn}{NLPN}{\abbr{n}onlinear \abbr{p}hase \abbr{n}oise}
\NewA{nc}{NC}{\abbr{n}etwork \abbr{c}oding}
\NewA{ofmd}{OFDM}{\abbr{o}rthogonal \abbr{f}requency-\abbr{d}ivision \abbr{m}ultiplexing}			
\NewA{dp}{DP}{\abbr{d}ual-\abbr{p}olarization}
\NewA{pam}{PAM}{\abbr{p}ulse \abbr{a}mplitude \abbr{m}odulation}
\NewA{pdf}{PDF}{\abbr{p}robability \abbr{d}ensity \abbr{f}unction}
\NewA{plnc}{PNC}{\abbr{p}hysical-layer \abbr{n}etwork \abbr{c}oding}
\NewA{psk}{PSK}{\abbr{p}hase-\abbr{s}hift \abbr{k}eying}
\NewA{pmd}{PMD}{\abbr{p}olarization \abbr{m}ode \abbr{d}ispersion}
\NewA{pm}{PM}{\abbr{p}olarization-\abbr{m}ultiplexed}
\NewA{pdm}{PDM}{\abbr{p}olarization-\abbr{d}ivision \abbr{m}ultiplexing}
\NewA{qam}{QAM}{\abbr{q}uadrature \abbr{a}mplitude \abbr{m}odulation}
\NewA{sqp}{SQP}{\abbr{s}equential \abbr{q}uadratic \abbr{p}rogramming}
\NewA{rd}{RD}{\abbr{r}adii \abbr{d}istribution}
\NewA{sep}{SEP}{\abbr{s}ymbol \abbr{e}rror \abbr{p}robability}
\NewA{ser}{SER}{\abbr{s}ymbol \abbr{e}rror \abbr{r}ate}
\NewA{si}{SI}{\abbr{s}ide \abbr{i}nformation}
\NewA{sp}{SP}{\abbr{s}ingle-\abbr{p}olarization}
\NewA{sanr}{SNR}{\abbr{s}ignal-to-(additive-)\abbr{n}oise \abbr{r}atio}
\NewA{snr}{SNR}{\abbr{s}ignal-to-\abbr{n}oise \abbr{r}atio}
\NewA{snlse}{sNLSE}{\abbr{s}tochastic \abbr{n}onlinear \abbr{S}chr\"odinger \abbr{e}quation}
\NewA{nlse}{NLSE}{\abbr{n}onlinear \abbr{S}chr\"odinger \abbr{e}quation}
\NewA{stwrc}{sTRC}{\abbr{s}eparated \abbr{t}wo-way \abbr{r}elay \abbr{c}hannel}
\NewA{stwtrc}{sTTRC}{\abbr{s}eparated \abbr{t}wo-way \abbr{t}wo-\abbr{r}elay \abbr{c}hannel}
\NewA{ts}{TS}{\abbr{t}wo-\abbr{s}tage}
\NewA{twrc}{TRC}{\abbr{t}wo-way \abbr{r}elay \abbr{c}hannel}
\NewA{twtrc}{TTRC}{\abbr{t}wo-way \abbr{t}wo-\abbr{r}elay \abbr{c}hannel}
\NewA{wdm}{WDM}{\abbr{w}avelength-\abbr{d}ivision \abbr{m}ultiplexing}
\NewA{vn}{VN}{variable node}
\NewA{cn}{CN}{constraint node}
\NewA{de}{DE}{density evolution}
\NewA{sc}{SC}{spatially-coupled}
\NewA{ldpc}{LDPC}{low-density parity-check}
\NewA{bp}{BP}{belief propagation}
\NewA{dm}{DM}{dispersion-managed}
\NewA{spm}{SPM}{self-phase modulation}
\NewA{xpm}{XPM}{cross-phase modulation}
\NewA{fwm}{FWM}{four-wave-mixing}
\NewA{ixpm}{IXPM}{intrachannel cross-phase modulation}
\NewA{ifwm}{IFWM}{intrachannel four-wave-mixing}
\NewA{ssfm}{SSFM}{split-step Fourier method}
\NewA{fec}{FEC}{forward error correction}
\NewA{psd}{PSD}{power spectral density}
\NewA{ook}{OOK}{on-off keying}
\NewA{smf}{SMF}{single-mode fiber}
\NewA{ssmf}{SSMF}{standard single-mode fiber}
\NewA{dbp}{DBP}{digital backpropagation}
\NewA{edfa}{EDFA}{erbium-doped fiber amplifier}
\NewA{ofdm}{OFDM}{orthogonal frequency division multiplexing}
\NewA{exit}{EXIT}{extrinsic information transfer}
\NewA{pexit}{P-EXIT}{protograph extrinsic information transfer}
\NewA{osnr}{OSNR}{optical signal-to-noise ratio}
\NewA{roadm}{ROADM}{reconfigurable optical add-drop multiplexer}
\NewA{rps}{RPS}{Raman pump station}
\NewA{mlse}{MLSE}{maximum likelihood sequence estimation}
\NewA{dcf}{DCF}{dispersion compensating fiber}
\NewA{tcm}{TCM}{trellis coded modulation}
\NewA{edc}{EDC}{electronic dispersion compensation}
\NewA{ldpcc}{LDPCC}{low-density parity-check convolutional}
\NewA{llr}{LLR}{log-likelihood ratio}
\NewA{ra}{RA}{repeat-accumulate}
\NewA{ira}{IRA}{irregular-repeat-accumulate}
\NewA{ara}{ARA}{accumulate-repeat-accumulate}
\NewA{mi}{MI}{mutual information}
\NewA{vdmm}{VDMM}{variable degree matched mapping}
\NewA{gmi}{GMI}{generalized mutual information}
\NewA{wd}{WD}{windowed decoder}
\NewA{gn}{GN}{Gaussian noise}
\NewA{hd}{HD}{hard-decision}
\NewA{hdd}{HDD}{hard-decision decoding}
\NewA{sd}{SD}{soft-decision}
\NewA{sdd}{SDD}{soft-decision decoding}
\NewA{emp}{EMP}{extrinsic message passing}
\NewA{imp}{IMP}{intrinsic message passing}
\NewA{gldpc}{GLDPC}{generalized low-density parity-check}
\NewA{scgldpc}{SC-GLDPC}{spatially-coupled generalized low-density parity-check}
\NewA{scldpc}{SC-LDPC}{spatially-coupled low-density parity-check}
\NewA{scfec}{SC-FEC}{spatially-coupled forward error correction}
\NewA{tbbc}{TBBC}{tightly-braided block code}
\NewA{gpc}{GPC}{generalized product code}
\NewA{otn}{OTN}{optical transport network}
\NewA{bch}{BCH}{Bose--Chaudhuri--Hocquenghem}
\NewA{rs}{RS}{Reed--Solomon}
\NewA{bbbc}{BBBC}{block-wise braided block code}
\NewA{hpc}{HPC}{half-product code}
\NewA{pc}{PC}{product code}
\NewA{rv}{RV}{random variable}
\NewA{mbp}{MBP}{multiple block permutator}
\NewA{cer}{CER}{codeword error probability}
\NewA{oh}{OH}{overhead}
\NewA{ids}{IDS}{informed dynamic scheduling}
\NewA{bdd}{BDD}{bounded-distance decoding}
\NewA{ncg}{NCG}{net coding gain}
\NewA{gwbp}{Galton--Watson branching process}{Galton--Watson branching process}

\newacronym[%
	longplural={binary erasure channels},%
	shortplural={BECs}%
]{bec}{BEC}{binary erasure channel}%

\newcommand{\BCH}{\gls{bch}\xspace}
\newcommand{\BEC}{\gls{bec}\xspace}
\newcommand{\BSC}{\gls{bsc}\xspace}
\newcommand{\DE}{\gls{de}\xspace}

\newcommand{\BDD}{\gls{bdd}\xspace}
\newcommand{\GPC}{\gls{gpc}\xspace}

\newcommand{\PCs}{\glspl{pc}\xspace}

\newcommand{\PC}{\gls{pc}\xspace}
\newcommand{\LDPC}{\gls{ldpc}\xspace}

\newcommand{\GPCs}{\glspl{gpc}\xspace}

\newcommand{\BER}{\gls{ber}\xspace}

\newcommand{\PP}{\gls{pp}\xspace}

\newcommand{\NCG}{\gls{ncg}\xspace}

\begin{document}

%\title{Approaching Miscorrection-free Performance\\ of Product and
%Generalized Product Codes\\ under Iterative Bounded-Distance Decoding}

\title{Approaching Miscorrection-free Performance\\ of Product and
Generalized Product Codes}

%\title{Miscorrection-free Decoding of Product\\ and
%Generalized Product Codes}

% use for special paper notices
%\IEEEspecialpapernotice{(Invited Paper)}
%\IEEEspecialpapernotice{(Draft, \today)}

\author{
	\IEEEauthorblockN{
	Christian Häger, \emph{Member, IEEE}
	and
	Henry D.~Pfister, \emph{Senior Member, IEEE}
	\thanks{Parts of this paper have been presented at the 2017 European
	Conference on Optical Communication (ECOC), Gothenburg, Sweden.}%
	\thanks{This work is part of a project that has received
	funding from the European Union's Horizon 2020 research and innovation
	programme under the Marie Sk\l{}odowska-Curie grant agreement
	No.~749798. The work was also supported in part by the National
	Science Foundation (NSF) under Grant No.~1609327. Any opinions,
	findings, recommendations, and conclusions expressed in this material
	are those of the authors and do not necessarily reflect the views of
	these sponsors.}%
	\thanks{C.~Häger is with the Department of Electrical Engineering,
	Chalmers University of Technology, SE-41296 Gothenburg, Sweden and the
	Department of Electrical and Computer Engineering, Duke University,
	Durham, NC 27708, USA (e-mail: christian.haeger@chalmers.se).
	H.~D.~Pfister is with the Department of Electrical and Computer
	Engineering, Duke University, Durham, NC 27708, USA (e-mail:
	henry.pfister@duke.edu).  
	}%
	}%
%	\IEEEauthorblockA{\IEEEauthorrefmark{1}%
%	Department of Electrical and Computer Engineering, Duke University,
%	Durham, North Carolina
%	}
%	\{ch303, henry.pfister\}@duke.edu 
}

\maketitle

%The algorithm relies on so-called anchor codewords to resolve
%inconsistencies across component codewords leading to significant
%performance improvements.

\begin{abstract}
	Product codes (PCs) protect a two-dimensional array of bits using
	short component codes. Assuming transmission over the binary
	symmetric channel, the decoding is commonly performed by iteratively
	applying bounded-distance decoding to the component codes. For this
	coding scheme, undetected errors in the component decoding---also
	known as miscorrections---significantly degrade the performance.  In
	this paper, we propose a novel iterative decoding algorithm for PCs
	which can detect and avoid most miscorrections.  The algorithm can
	also be used to decode many recently proposed classes of generalized
	PCs such as staircase, braided, and half-product codes. Depending on
	the component code parameters, our algorithm significantly
	outperforms the conventional iterative decoding method. As an
	example, for double-error-correcting Bose--Chaudhuri--Hocquenghem
	component codes, the net coding gain can be increased by up to
	$0.4\,$dB.  Moreover, the error floor can be lowered by orders of
	magnitude, up to the point where the decoder performs virtually
	identical to a genie-aided decoder that avoids all miscorrections.
	We also discuss post-processing techniques that can be used to
	reduce the error floor even further.  
\end{abstract}

% Redefine all acronyms that have been defined in the introduction
\glsresetall

\section{Introduction}

A \PC is the set of all $n \times n$ arrays where each row and column
in the array is a codeword in some linear component code $\mathcal{C}$
of length $n$ \cite{Elias1954}. Recently, a wide variety of related
code constructions have been proposed, e.g., braided codes
\cite{Feltstrom2009}, half-product codes \cite{Justesen2010,
Justesen2011}, continuously-interleaved codes \cite{Scholten2010},
half-braided codes \cite{Justesen2011, Pfister2015}, and staircase
codes \cite{Smith2012a}. All of these code classes have Tanner graph
representations that consist exclusively of degree-2 variable nodes,
i.e., each bit is protected by two component codes. We use the term
\GPCs to refer to such codes.

The component codes of a \GPC typically correspond to Reed--Solomon or
\BCH codes, which can be efficiently decoded via algebraic \BDD. The
overall \GPC is then decoded by iteratively applying \BDD to the
component codes. This iterative coding scheme dates back to 1968
\cite{Abramson1968} and has been shown to offer excellent performance
in practice. In particular for the \BSC at high code rates, iterative
decoding of \GPCs with binary \BCH component codes can achieve
performance close to the channel capacity
\cite{Smith2012a,Jian2014,Jian2017}. Moreover, the decoder data flow
can be orders of magnitude lower than for comparable \LDPC codes under
message-passing decoding \cite{Smith2012a}. This facilitates decoder
throughputs of tens or even hundreds of Gigabits per second. Indeed,
\GPCs are popular choices for high-bit-rate applications with limited
soft information such as regional/metro optical transport networks
\cite{Smith2012a,Farhoodfar2011, Scholten2010, Jian2014, Zhang2014,
Haeger2015ofc, Justesen2010, Justesen2011, Pfister2015, Haeger2017tit,
Haeger2016ofc, Condo2017}.  Besides data transmission, \GPCs are also
used in storage applications
\cite{Chang2001,Kim2007,Vo2011,Yang2012,Emmadi2015}. 

%such as DVDs \cite{Chang2001}, embedded SRAM caches \cite{Kim2007},
%high-density magnetic recording \cite{Vo2011}, and flash memory
%\cite{Yang2012, Emmadi2015}. 

%A well-known problem associated with the iterative decoding of \GPCs
%on the \BSC are undetected decoding errors, also known as
%miscorrections, that arise during \BDD. 

For \GPCs over the \BSC, undetected errors in the component
decoding---also known as miscorrections---significantly degrade the
performance of iterative decoding. In particular, let $\vect{r} =
\vect{c} + \evec$, where $\vect{c}, \evec \in \{0,1\}^n$ denote a
component codeword and random error vector, respectively. For a
$t$-error-correcting component code $\mathcal{C}$, BDD yields the
correct codeword $\vect{c} \in \mathcal{C}$ if and only if
$d_\text{H}(\vect{r}, \vect{c}) = w_\text{H}(\evec) \leq t$, where
$d_\text{H}$ and $w_\text{H}$ denote Hamming distance and weight,
respectively. On the other hand, if $w_\text{H}(\evec) > t$, the
decoding either fails or there exists another codeword $\vect{c}' \in
\mathcal{C}$ such that $d_\text{H}(\vect{r},\vect{c}') \leq t$. In the
latter case, we say that a miscorrection occurs, in the sense that
\BDD is technically successful but the decoded codeword $\vect{c}'$ is
not the correct one. Miscorrections are highly undesirable because
they introduce additional errors (on top of channel errors) into the
iterative decoding process. Moreover, from a theoretical perspective,
miscorrections are notoriously difficult to analyze in an iterative
decoding scheme \cite{Schwartz2005, Justesen2007, Justesen2011,
Smith2012a, Haeger2017tit, Zhang2015, Truhachev2016, Holzbaur2017}. In
fact, despite the widespread use in practice and to the best of our
knowledge, no rigorous analytical results exist characterizing the
finite-length performance of \GPCs under iterative \BDD over the \BSC. 

For specific code proposals in practical systems, the miscorrection
problem is typically addressed by appropriately modifying the
component code that is used to construct the \GPC, see, e.g.,
\cite{Justesen2011, Smith2012a, Jian2014, Condo2017}. In particular,
for binary $t$-error-correcting BCH codes, it is known that
miscorrections occur approximately with probability $1/t!$
\cite{McEliece1986, Justesen2011}. In order to reduce this
probability, one may employ a subcode of the original code
\cite{Justesen2011, Smith2012a}, extend the code \cite{Condo2017},
and/or apply code shortening \cite{Smith2012a, Condo2017}. On the
other hand, such modifications invariably lead to a code rate reduction.
Moreover, even with a modified component code, miscorrections can still
have a significant effect on the performance. 

The main contribution in this paper is a novel iterative decoding
algorithm for \GPCs which can detect and avoid most miscorrections.
The algorithm relies on so-called anchor codewords to resolve
inconsistencies across component codewords. This leads to significant
performance improvements, in particular when $t$ is small (which is
typically the case in practice). As an example, for $t=2$, the
algorithm can improve the net coding gain by roughly $0.4\,$dB.
Moreover, the error floor can be lowered by almost two orders of
magnitude, up to the point where the performance is virtually
identical to that of a miscorrection-free genie-aided decoder.

Error-floor improvements are particularly important for applications
with stringent reliability constraints such as optical transport
networks. We therefore also discuss the application of \PP techniques
\cite{Sridharan2003, Jian2014, Emmadi2015, Condo2016b,
Mittelholzer2016, Holzbaur2017}, which can be combined with the
proposed algorithm to reduce the error floor even further. 

Decoder modifications that target miscorrections have been proposed
before in the literature. Usually these modifications are minor and
they are tailored to a specific \GPC. As an example, staircase codes
\cite{Smith2012a} can be seen as a convolutional-like (or
spatially-coupled) version of \PCs. The associated code array consists
of an infinite number of square blocks that are arranged to look like a
staircase \cite[Fig.~4]{Smith2012a}. Decoding is facilitated by using
a sliding window which comprises only a finite number of blocks.  In
order to reduce miscorrections, one may reject certain bit flips from
component codewords that are associated with the newest (most
unreliable) block from the channel \cite[p.~59]{Smith2011}. However,
simulations suggest that the performance gains using this approach are
limited. On the other hand, the proposed anchor-based decoding can
closely approach miscorrection-free performance when applied to
staircase codes \cite{Haeger2017ecoc}.

%and it is neither mentioned in the original journal paper
%\cite{Smith2012a} nor in the patent application \cite{Cortina2011}. 

In \cite{Justesen2011}, a decoder modification for \PCs is suggested
based on the observation that the miscorrection probability is reduced
by a factor of $n$ if only $t-1$ errors are corrected for a
$t$-error-correcting component code. For large $n$, this is
significant and the author thus proposes to only correct $t-1$ errors
in the first iteration of iterative \BDD. We will see later that this
indeed gives some notable performance improvements.  This trick can
also be easily combined with the proposed algorithm.  

The work in this paper is inspired by another comment made in
\cite{Justesen2011} where it is mentioned that for \PCs it may be
desirable to ``take special actions in case of conflicts'' caused by
miscorrections. In particular, it is suggested to use the number of
conflicts for a particular component code as an indicator for the
reliability of the component code. Besides this suggestion, no further
details or results are provided. Our algorithm builds upon this idea
and we develop a systematic approach that exploits component code
conflicts and can be applied to an arbitrary \GPC. 

%The idea is to ``initially count the number of bits involved in
%conflicts in any particular component code and then to give a code
%with a larger number of conflicts a lower weight in the decisions''.
%However, it is not exactly clear how to translate this comment into a
%tangible algorithm since no further details are provided. Besides this
%comment, no further details are provided. 

The remainder of the paper is structured as follows. In
Sec.~\ref{sec:product_codes}, we review \PCs and the conventional
iterative decoding scheme. We also give a brief overview of
theoretical methods that have been proposed to analytically predict
the performance. The proposed anchor-based decoding algorithm is
described in Sec.~\ref{sec:anchor_based_decoding}. In
Sec.~\ref{sec:results}, simulation results are presented and discussed
for various component code parameters. \PP is discussed in
Sec.~\ref{sec:post_processing}. We discuss the complexity and some
implementation issues for the proposed algorithm in
Sec.~\ref{sec:complexity}. Finally, the paper is concluded in
Sec.~\ref{sec:conclusion}. 

\section{Product Codes and Iterative Decoding}
\label{sec:product_codes}

This paper focuses on \PCs, which we believe many readers are familiar
with. Other classes of \GPCs are discussed separately in
Sec.~\ref{sec:gpcs}. 

\subsection{Product Codes}

Let $\mat{H} \in \GF{2}^{(n-k) \times n}$ be the parity-check matrix
of a binary linear $(n,k,\dmin)$ code $\mathcal{C}$, where $n$, $k$,
and $\dmin$ are the code length, dimension, and minimum distance,
respectively. A \PC based on $\mathcal{C}$ is defined as
\begin{align}
	\label{eq:pc_definition}
	\mathcal{P}(\mathcal{C}) \define \{\mat{X} \in \GF{2}^{n \times n}
	\,|\, \mat{H}\mat{X} = \mat{0}, \mat{X}\mat{H}^\transpose = \mat{0}
	\}.
\end{align}
It can be shown that $\mathcal{P}(\mathcal{C})$ is a linear
$(n^2,k^2,\dmin^2)$ code.  
The codewords $\mat{X}$ can be represented as two-dimensional arrays.
The two conditions in \eqref{eq:pc_definition} enforce that the rows
and columns in the array are valid codewords in $\mathcal{C}$. 

We use a pair $(i,j)$ to identify a particular component codeword. The
first parameter $i \in \{1,2\}$ refers to the codeword type which can
be either a row $(i=1)$ or a column $(i=2)$. The second parameter $j
\in [n]$ enumerates the codewords of a given type. 

\begin{figure}[t]
	\centering
		\includegraphics{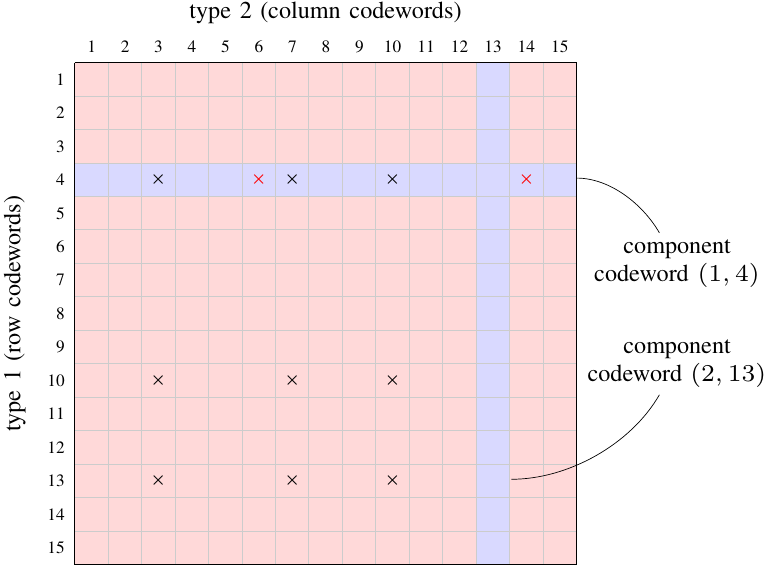}
	\caption{\PC array for a length-$15$ component code. Two particular
	component codewords are highlighted. The black crosses correspond to
	a minimal-size stopping set when $t=2$. Red crosses indicate a
	miscorrection, see Example~\ref{ex:miscorrection}.}
	\label{fig:product_code_array}
\end{figure}

\begin{example}
	The code array for a \PC where the component code has length $n=15$
	is shown in Fig.~\ref{fig:product_code_array}. The component
	codewords $(1,4)$ and $(2,13)$ are highlighted in blue. \demo
\end{example}

For \PCs, the coded bits can be identified by their two coordinates
within the \PC array. However, this way of specifying bits does not
generalize well to other classes of \GPCs because the associated code
array may have a different shape (or there may not exist an array
representation at all). In order to keep the notation general, we
therefore use the convention that a coded bit is specified by two
component codewords $(i,j)$ and $(k,l)$, i.e., four parameters
$(i,j,k,l)$ in total. 

\begin{example}
	The two highlighted component codewords in
	Fig.~\ref{fig:product_code_array} intersect at the bit corresponding
	to the tuple $(1,4,2,13)$ or $(2,13,1,4)$.  \demo
\end{example}

%\begin{remark}
%	\PCs have a deterministic Tanner graph representation that resembles
%	a complete bipartite graph. There exist two types of $n$ \CNs each
%	(i.e., $2n$ \CNs in total), corresponding to the row and column
%	codes, respectively. The Tanner graph is such that each \CN of type
%	1 is connected to each \CN of type 2 through a \VN. This gives rise
%	to a total of $n^2$ \VNs, where each \VN corresponds to one bit in
%	the code array. For a graphical representation, see, e.g.,
%	\cite[Fig.~3]{Tanner1981}. 
%\end{remark}

\subsection{BCH Component Codes}
\label{sec:bch}

We use binary $t$-error-correcting \BCH codes as component
codes, as well as their singly- and doubly-extended versions. Recall
that a singly-extended \BCH code is obtained through an additional
parity bit, formed by adding (modulo 2) all coded bits
$c_1, c_2, \dots, c_{2^\nu-1}$ of the \BCH code, where $\nu$ is the
Galois field extension degree. On the other hand, a doubly-extended
\BCH code has two additional parity bits, denoted by $c_{2^\nu}$ and
$c_{2^\nu + 1}$, such that
\begin{align}
	c_1 + c_3 + \dots + c_{2^{\nu}-1} + c_{2^{\nu}+1} &= 0, \\
	c_2 + c_4 + \dots + c_{2^{\nu}-2} + c_{2^{\nu}} &= 0,
\end{align}
i.e., the parity bits perform checks separately on odd and even bit
positions. The overall component code has length $n = 2^{\nu}-1+e$,
where $e \in \{0,1,2\}$ indicates either no ($e=0$), single ($e=1$),
or double ($e=2$) extension. In all three cases, the guaranteed code
dimension is $k = 2^\nu-1- \nu t$. For $e=0$, the guaranteed minimum
distance is $d_\text{min} = 2t+1$. This is increased to $d_\text{min}
= 2t+2$ for $e \in \{1,2\}$. We use a triple $(\nu, t, e)$ to denote
all \BCH code parameters. 

\subsection{Bounded-Distance Decoding and Miscorrections}
\label{sec:bdd}

Consider the transmission of a component codeword $\vect{c} \in
\mathcal{C}$ over the \BSC with crossover probability $p$. The
error vector introduced by the channel is denoted by $\evec$, i.e.,
the components of $\evec$ are \IID Bernoulli($p$) random variables.
Applying \BDD to the received vector $\vect{r} = \vect{c} + \evec$
results in
\begin{align}
	\label{eq:bdd}
	\text{BDD}(\vect{r}) = \begin{cases}
		\vect{c} \quad &\text{if }d_\text{H}(\vect{r},\vect{c})
		= w_\text{H}(\evec) \leq t, \\
		\vect{c}' \in \mathcal{C} \quad &\text{if }w_\text{H}(\evec) > t
		\text{ and } d_\text{H}(\vect{r},\vect{c}') \leq t,\\
		\text{\tt FAIL} &\text{otherwise}.
	\end{cases}
\end{align}
In practice, \BDD is implemented by first computing the syndrome
$\vect{s}^\transpose = \mat{H} \vect{r}^\transpose = \mat{H}
\vect{n}^\transpose \in \GF{2}^{n-k}$. Each of the $2^{n-k}$ possible
syndromes is then associated with either an estimated error vector
$\hat{\evec}$, where $w_\text{H}(\hat{\evec}) \leq t$, or a decoding
failure. In the first case, the decoded output is computed as
$\vect{r} + \hat{\evec}$. 

The second case in \eqref{eq:bdd} corresponds to an undetected error
or miscorrection. 

\begin{example}
	\label{ex:miscorrection}
	Consider the component codeword $(1,4)$	in
	Fig.~\ref{fig:product_code_array} and assume that the all-zero
	codeword $\vect{c} = \vect{0}$ is transmitted. The black crosses
	represent bit positions which are received in error, i.e., $n_i = 1$
	for $i \in \{3,7,10\}$ and $n_i = 0$ elsewhere. For a component code
	with $t=2$ and $e=0$, we have $\dmin = 2t+1 = 5$, i.e., there exists
	at least one codeword $\vect{c}' \in \mathcal{C}$ with Hamming
	weight 5. Assume we have $\vect{c}' \in \mathcal{C}$ with $c_i' = 1$
	for $i \in \{3,6,7,10,14\}$ and $c_i' = 0$ elsewhere. Applying \BDD
	to $\vect{r} = \vect{c} + \evec$ then introduces two additional
	errors at bit positions 6 and 14. This is shown by the red crosses
	in Fig.~\ref{fig:product_code_array}. \demo
\end{example}

Code extension reduces the probability of miscorrecting, at the
expense of a slightly increased code length (and hence a small rate
loss). To see this, let $e=0$ and assume that $w_\text{H}(\evec) > t$.
Consider now the decoding of a random syndrome $\vect{s}$ where the
components in $\vect{s}$ are \IID Bernoulli($0.5$). In that case, the
probability of miscorrecting is simply the ratio of the number of
decodable syndromes and the total number of syndromes, i.e., 
\begin{align}
	\label{eq:miscorrection_probability}
	\frac{\sum_{i=0}^t \binom{n}{i} }{2^{n-k}} = 
	\frac{\sum_{i=0}^t \binom{n}{i} }{2^{\nu t}} \approx
	\frac{\frac{1}{t!} n^t }{n^t} = \frac{1}{t!}.
\end{align}
For $e = 1$ and $e = 2$, the total number of syndromes is increased to
$2^{\nu t + e}$ and the miscorrection probability is thus reduced by a
factor of 1/2 and 1/4, respectively. We note that this reasoning can
be made precise, see, e.g., \cite{McEliece1986, Justesen2011}. 

\begin{example}
	Consider the same scenario as in Example \ref{ex:miscorrection}. If
	we use the singly-extended component code, the minimum distance is
	increased to $\dmin = 2t+2 = 6$. Assuming that there is no
	additional error in the last bit position, i.e., $n_{16} = 0$, the
	miscorrection illustrated in Example \ref{ex:miscorrection} can be
	detected because the parity-check equation involving the additional
	parity bit is not satisfied. \demo
\end{example}

\begin{remark}
	As an alternative to extending the code, one may employ a subcode of
	the original BCH code. For example, the singly-extended BCH code
	behaves similarly to the even-weight subcode of the BCH code, which is
	obtained by multiplying its generator polynomial by $(1+x)$. The
	subcode where odd and even coded bits separately sum to zero is
	obtained by multiplying the generator polynomial by $(1+x^2)$.
	Subcodes have a reduced code dimension $k$ and hence lead to a
	similar rate loss as the code extension. 
\end{remark}

\subsection{Iterative Bounded-Distance Decoding}
\label{sec:performance}

We now consider the transmission of a codeword $\mat{X} \in
\mathcal{P}(\mathcal{C})$ over the \BSC with crossover probability
$p$. The conventional iterative decoding procedure consists of
applying \BDD first to all row component codewords and then to all
column component codewords. This is repeated $\ell$ times or until a
valid codeword in $\mathcal{P}(\mathcal{C})$ is found. Pseudocode for
the iterative \BDD is given in Algorithm \ref{alg:iterative_bdd}.
A stopping-criterion is omitted for readability purposes. 

\newcommand{\algtext}[1]{\text{\sf\scriptsize #1}}
\setlength{\textfloatsep}{10pt}
\begin{algorithm}[t]
	%\SetAlgorithmName{MegaAlgorithm}{}
	\small
	\DontPrintSemicolon
	\SetKw{ShortFor}{for}
	\SetKw{Break}{break}
	\SetKw{MyWhile}{while}
	\SetKw{MyIf}{if}
	\SetKw{MySet}{set}
	\SetKw{MyElse}{else}
	\SetKw{MyCompute}{compute}
	\SetKw{KwEach}{each}
	\SetKw{KwAnd}{and}
	%\SetKwData{maxIter}{maxIter}
	%\SetKwFunction{VNupdate}{VNupdate}

	%$k \leftarrow 0$\;
	\For{$l = 1, 2, \dots, \ell$}{
		\For{$i = 1, 2$}{
			\For{$j = 1, 2, \dots, n$}{
				apply BDD to component codeword $(i,j)$
			}
		}
	}
	%	output decision for $\mat{B}_k$ and shift window\;
	%	$k \leftarrow k + 1$\;
	\caption{ {\small Iterative BDD of product codes} }
	\label{alg:iterative_bdd}
\end{algorithm}
%\DecMargin{1em}

In order to analyze the \BER of \PCs under iterative \BDD, the
prevailing approach in the literature is to assume that no
miscorrections occur in the BDD of the component codes, see, e.g.,
\cite{Schwartz2005, Justesen2007, Justesen2011, Haeger2017tit,
Zhang2015}. To that end, we define 
\begin{align}
	\label{eq:idealized_bdd}
	\text{BDD}'(\vect{r}) = \begin{cases}
		\vect{c} \quad &\text{if }d_\text{H}(\vect{r},\vect{c})
		= w_\text{H}(\evec) \leq t, \\
		\text{\tt FAIL} &\text{otherwise},
	\end{cases}
\end{align}
which can be seen as an idealized version of \BDD where a genie
prevents miscorrections. Conceptually, this is similar to assuming
transmission over the \BEC instead of the \BSC \cite{Schwartz2005}. 

Using \eqref{eq:idealized_bdd} instead of \eqref{eq:bdd}, a decoding
failure for a \PC is related to the existence of a so-called core in an \ErdRen
random graph \cite{Justesen2011, Justesen2007}. This connection can be
used to rigorously analyze the asymptotic performance as $n \to
\infty$ using \DE \cite{Haeger2017tit} \cite{Justesen2011}
\cite{Justesen2007}. Moreover, the error floor can be estimated by
enumerating stopping sets, also known as stall patterns.
A stopping set is a subset of bit positions such that every
component codeword with at least one bit in the set must contain at
least $t+1$ bits in the set. For \PCs, a minimal-size stopping set involves
$t+1$ row codewords and $t+1$ column codewords and has size $\smin =
(t+1)^2$. For example, the black crosses shown in
Fig.~\ref{fig:product_code_array} form such a stopping set when $t=2$.
If we consider only stopping sets of minimal size, the \BER can be
approximated as
\begin{align}
	\label{eq:pc_error_floor}
	\text{BER} \approx \frac{\smin}{n^2} M p^{\smin},
\end{align}
for sufficiently small $p$, where $M = \binom{n}{t+1}^2$ is the total
number of possible minimal-size stopping sets, also referred to as the
stopping set's multiplicity. Unfortunately, if miscorrections are
taken into account, \DE and the error floor analysis are nonrigorous
and become inaccurate. 

\begin{figure}[t]
	\begin{center}
		\includegraphics{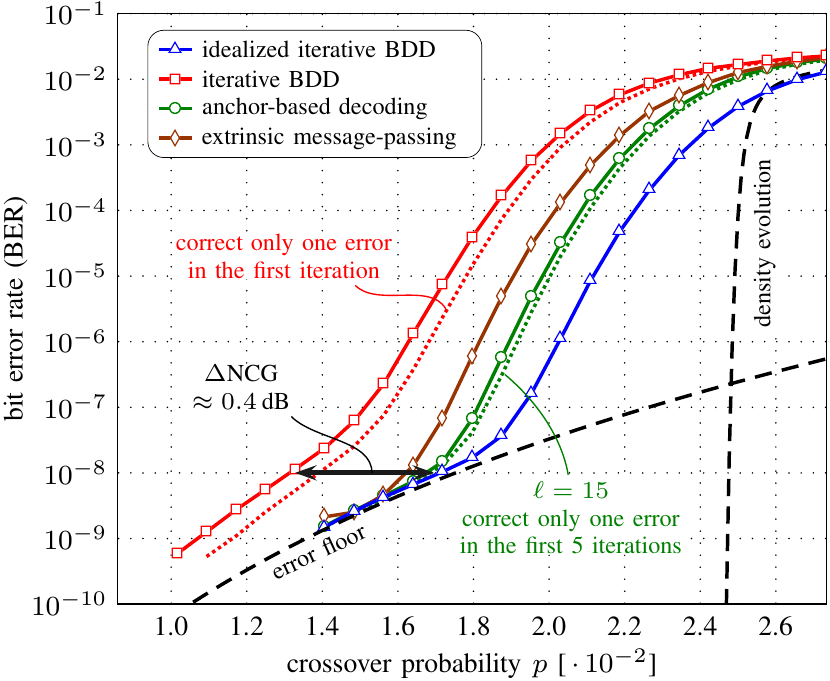}
	\end{center}
	\caption{Simulation results using different
	decoding schemes with $\ell = 10$ iterations. The component code is
	a BCH code with parameters $(7,2,1)$, i.e., 
	an extended double-error-correcting BCH code with length
	$n = 2^7 = 128$. }
	\label{fig:performance_gap}
\end{figure}

\begin{example} 
	\label{ex:performance_gap}
	Consider a \BCH code $\mathcal{C}$ with parameters $(7,2,1)$. The
	resulting \PC $\mathcal{P}(\mathcal{C})$ has length $n^2 = 128^2 =
	16384$ and code rate $R = k^2/n^2 \approx 0.78$. For $\ell = 10$
	decoding iterations, the outcome of \DE (see Appendix \ref{app:de}
	for details) and the error floor analysis via
	\eqref{eq:pc_error_floor} are shown in
	Fig.~\ref{fig:performance_gap} by the dashed black lines.  The
	analysis can be verified by performing idealized iterative \BDD
	using \eqref{eq:idealized_bdd}. The results are shown by the blue
	line (triangles). However, the actual \BER with true BDD
	\eqref{eq:bdd} deviates significantly from the idealized decoding,
	as shown by the red line (squares). The \BER can be moderately
	improved by treating the component codes as single-error-correcting
	in the first iteration, as suggested in \cite{Justesen2011}.  This
	is shown by the red dotted line. \demo
\end{example}
%For convenience, we have included a self-contained description of \DE
%for \GPCs in Appendix \ref{app:de}.

There exist several approaches to quantify the performance loss due to
miscorrections. In terms of the error floor, the authors in
\cite{Smith2012a} derive an expression similar to
\eqref{eq:pc_error_floor} for staircase codes. To account for
miscorrections, this expression is modified by introducing a heuristic
parameter, whose value has to be estimated using Monte--Carlo
simulations. In terms of asymptotic performance, the authors in
\cite{Truhachev2016} recently proposed a novel approach to analyze a
\GPC ensemble that is structurally related to staircase codes.  The
presented method is shown to give more accurate asymptotic predictions
compared to the case where miscorrections are ignored. 

Rather than analyzing the effect of miscorrections, the
approach taken in this paper is to try to avoid them by modifying the
decoding. In the next section, we give a detailed description of the
proposed decoding algorithm. Its \BER performance for the code
parameters considered in Example \ref{ex:performance_gap} is shown in
Fig.~\ref{fig:performance_gap} by the green line (circles). The
results are discussed in more detail in Sec.~\ref{sec:results} below.

%A thorough comparison and discussion of the algorithm with
%respect to the state-of-the art is presented in
%Sec.~\ref{sec:results}.

%In fact, the main motivation behind this work stems from the
%difficulty of theoretically analyzing the performance of iterative
%\BDD due to the effect of miscorrections. Rather than analyzing the ,
%our approach is to try to avoid them. 

\begin{remark}
	\label{rmk:emp}
	Iterative \BDD can be interpreted as a message-passing algorithm
	with binary ``hard-decision'' messages. The corresponding
	message-passing rule is \emph{intrinsic}, in the sense that the
	outgoing message along some edge depends on the incoming message
	along the same edge. In \cite{Jian2017}, the authors propose an
	\emph{extrinsic} message-passing algorithm based on \BDD. The \BER
	for this algorithm when applied to the \PC in Example
	\ref{ex:performance_gap} is shown in Fig.~\ref{fig:performance_gap}
	by the brown line (diamonds). Similar to the proposed algorithm,
	extrinsic message-passing provides significant performance
	improvements over iterative \BDD. However, it is known that the
	decoder data-flow and storage requirements can be dramatically
	increased  for message-passing decoding compared to iterative \BDD
	\cite{Smith2012a}. One reason for this is that iterative \BDD can
	leverage a syndrome compression effect by operating entirely in the
	syndrome domain. We show in Sec.~\ref{sec:complexity} that this
	effect also applies to the proposed algorithm. For extrinsic
	message-passing, it is an open question if an efficient syndrome
	domain implementation is possible.  Due to this, we do not consider
	the extrinsic message-passing further in this paper. 
%	This algorithm can be rigorously analyzed via \DE on the \BSC
%	including miscorrections. This analysis, however, does not apply to
%	deterministic \GPCs such as \PCs, but only to properly defined code
%	ensembles.\footnote{The \DE in \cite{Justesen2007,
%	Justesen2011,Haeger2017tit} on the other hand does apply to
%	deterministic codes, but it is only rigorous on the \BEC or the \BSC
%	with idealized \BDD.} Nonetheless, the algorithm can still be used
%	to decode deterministic \GPCs. 
\end{remark}

\section{Anchor-based Decoding}
\label{sec:anchor_based_decoding}

In the previous section, we have seen that there exists a significant
performance gap between iterative \BDD and idealized iterative \BDD
where a genie prevents miscorrections. Our goal is to close this gap.
In order to do so, the key observation we exploit is that
miscorrections lead to inconsistencies (or conflicts) across component
codewords. In particular, two component codes that protect the same
bit may disagree on its value. In this section, we show how these
inconsistencies can be used to (a) reliably prevent miscorrections and
(b) identify miscorrected codewords in order to revert their decoding
decisions. 

\subsection{Preliminaries}

The proposed decoding algorithm relies on so-called anchor codewords
which have presumably been decoded without miscorrections. Roughly
speaking, we want to ensure that bit flips do not lead to
inconsistencies with anchor codewords. Consequently, decoding
decisions from codewords that are in conflict with anchors are not
applied. On the other hand, some anchor codewords may actually be
miscorrected. We therefore allow for the decoding decisions of anchors
to be overturned if too many other component codewords are in conflict
with a particular anchor. In order to make this more precise, we start
by introducing some additional concepts and notation in this
subsection. 

First, consider the \BDD of a single component codeword.  We
explicitly regard this component decoding as a two-step process. In
the first step, the actual decoding is performed and the outcome is
either an estimated error vector $\hat{\vect{n}}$ or a decoding
failure. In the second step, error-correction is performed by flipping
the bits corresponding to the error locations. These two steps are
separated in order to perform consistency checks (described below).
These checks are used to determine if the error-correction step should
be applied. 

It is more convenient to specify the estimated error
vector $\hat{\vect{n}}$ in terms of a set of error locations.  For
component codeword $(i,j)$, this set is denoted by
$\mathcal{E}_{i,j}$, where $|\mathcal{E}_{i,j}| \leq t$. The set
comprises those component codewords that are affected by the bit flips
implied by $\hat{\vect{n}}$. 

\begin{example}
	Consider again the scenario described in Example
	\ref{ex:miscorrection}, where the component codeword $(1,4)$ shown
	in Fig.~\ref{fig:product_code_array} miscorrects with an estimated
	error vector $\hat{\vect{n}}$ such that $\hat{n}_i = 1$ for $i \in
	\{6, 14\}$ and $\hat{n}_i = 0$ elsewhere. The corresponding set of
	error locations is given by $\mathcal{E}_{1,4} = \{(2,6), (2,14)\}$.
	\demo
\end{example}

\begin{remark}
	It may seem more natural to define $\mathcal{E}_{i,j}$ in terms of
	the bit positions of the BCH code, e.g., $\mathcal{E}_{1,4} = \{6,
	14\}$ in the previous example. However, defining $\mathcal{E}_{i,j}$
	in terms of the affected component codewords leads to a more
	succinct description of the proposed algorithm. Moreover, this
	definition also generalizes more easily to other classes of \GPCs,
	see Sec.~\ref{sec:gpcs} below. 
\end{remark}

Furthermore, we use a set $\mathcal{L}_{i,j}$ for each component
codeword that comprises other component codewords that are in conflict
with codeword $(i,j)$ due to miscorrections. Lastly, each component
codeword has an associated status to signify its current state. The
status values range from $0$ to $3$ with the following meaning:

\begin{itemize}
	\item 0: anchor codeword
	\item 1: eligible for BDD 
	\item 2: BDD failed in last iteration
	\item 3: frozen codeword
\end{itemize}
The precise use of the status and the transition rules between
different status values are described in the following.

\subsection{Main Algorithm Routine}

The algorithm is initialized by setting the status of all component
codewords to 1.  We then iterate $\ell$ times over the component
codewords in the same fashion as in Algorithm \ref{alg:iterative_bdd},
but replacing line 4 with lines 1--19 in Algorithm
\ref{alg:main_routine}. Algorithm \ref{alg:main_routine} represents
the main routine of the proposed anchor-based decoding. It can be
divided into 4 steps which are described in the following. 

\newcommand{\rightcomment}[1]{\tcc*[r]{#1}}
\newcommand\mycommfont[1]{\tiny\ttfamily\textcolor{blue}{#1}}
\SetCommentSty{mycommfont}

\IncMargin{1em}
%\restylealgo{boxed}
%\linesnumbered
\begin{algorithm}[t]
	\footnotesize
	\DontPrintSemicolon
	\SetKw{ShortFor}{for}
	\SetKw{Break}{break}
	\SetKw{MyWhile}{while}
	\SetKw{MyIf}{if}
	\SetKw{MySet}{set}
	\SetKw{MyElse}{else}
	\SetKw{MyCompute}{compute}
	\SetKw{KwEach}{each}
	\SetKw{KwAnd}{and}
	%\SetKwData{maxIter}{maxIter}
	\SetKwFunction{errProb}{error probability}
	%\EmptyLine
%	\KwIn{error graph $\mathcal{G}$} 
%	\KwOut{decoded error graph}
	%\BlankLine

	\If{$(i,j).\algtext{status} = 1$}{%
		$R \leftarrow (i,j).\algtext{decode}$ \rightcomment{R indicates
		success (1) or failure (0)}%
		%$\mathcal{B}.\algtext{clear}()$ \rightcomment{reset backtracking list}%
		\eIf{$R = 1$}{%
			\For(\tcc*[f]{consistency checks}){\KwEach $(k,l) \in \mathcal{E}_{i,j}$ }{%
				\If(\tcc*[f]{conflict with anchor}){$(k,l).\algtext{status} = 0$}{%
					\eIf{$|\mathcal{L}_{k,l}| \geq \delta$}{%
						add $(k,l)$ to $\mathcal{B}$ \rightcomment{mark for backtracking}%
						%$\mathcal{B}.\algtext{push}(k,l)$
					}{%
						$(i,j).\algtext{status} \leftarrow 3$ \rightcomment{freeze codeword}
						add $(k,l)$ to $\mathcal{L}_{i,j}$ \rightcomment{save the conflict}
						add $(i,j)$ to $\mathcal{L}_{k,l}$ \rightcomment{save the conflict}
	%					$(\mathcal{L}_{i,j}).\algtext{push}(k,l)$ \rightcomment{save the conflict}
	%					$(\mathcal{L}_{k,l}).\algtext{push}(i,j)$ \rightcomment{save the conflict}
					}%
				}%
			}%
			\If(\tcc*[f]{if not frozen}){$(i,j).\algtext{status} = 1$}{
				\For{\KwEach $(k,l) \in \mathcal{E}_{i,j}$}{
					error-correction for $(i,j,k,l)$ \rightcomment{see
					Alg.~\ref{alg:error_correction}}
				}%
				$(i.j).\algtext{status} \leftarrow 0$ \rightcomment{codeword becomes an anchor}
				\For{\KwEach $(k,l) \in \mathcal{B}$ }{
					backtrack anchor $(k,l)$ \rightcomment{see Alg.~\ref{alg:backtracking}}
				}%
			}%
		}{%
		$(i,j).\algtext{status} \leftarrow 2$ \rightcomment{indicate decoding failure}
		}%
	}%
	%
	% old code:
	%
%	\If(\tcc*[f]{codeword is decodable}){$(i,j).\algtext{status} = 1$}{%
%		%$\mathcal{B}.\algtext{clear}()$ \rightcomment{reset backtracking list}%
%		\For(\tcc*[f]{consistency checks}){\KwEach $(k,l) \in \mathcal{E}_{i,j}$ }{%
%			\If(\tcc*[f]{conflict with anchor}){$(k,l).\algtext{status} = 0$}{%
%				\eIf{$|\mathcal{L}_{k,l}| \geq \delta$}{%
%					add $(k,l)$ to $\mathcal{B}$ \rightcomment{mark for backtracking}%
%					%$\mathcal{B}.\algtext{push}(k,l)$
%				}{%
%					$(i,j).\algtext{status} \leftarrow 3$ \rightcomment{freeze component codeword}
%					add $(k,l)$ to $\mathcal{L}_{i,j}$ \rightcomment{save the conflict}
%					add $(i,j)$ to $\mathcal{L}_{k,l}$ \rightcomment{save the conflict}
%%					$(\mathcal{L}_{i,j}).\algtext{push}(k,l)$ \rightcomment{save the conflict}
%%					$(\mathcal{L}_{k,l}).\algtext{push}(i,j)$ \rightcomment{save the conflict}
%				}%
%			}%
%		}%
%		\If(\tcc*[f]{if not frozen}){$(i,j).\algtext{status} = 1$}{
%			\For{\KwEach $(k,l) \in \mathcal{B}$ }{
%				backtrack anchor $(k,l)$ \rightcomment{see Alg.~\ref{alg:backtracking}}
%			}%
%			\For{\KwEach $(k,l) \in \mathcal{E}_{i,j}$}{
%				error-correction step for $(i,j,k,l)$ \rightcomment{see
%				Alg.~\ref{alg:error_correction}}
%			}%
%			$(i.j).\algtext{status} \leftarrow 0$ \rightcomment{component codeword becomes an anchor}
%		}%
%	}%
	\caption{\small Main routine of anchor-based decoding}
	\label{alg:main_routine}
\end{algorithm}
\DecMargin{1em}

\subsubsection*{Step 1 (Lines 1--3)}

If the component codeword is eligible for \BDD, i..e, its status is 1,
we proceed to decode the component codeword. If the decoding is
successful, we proceed to the next step, otherwise, the codeword
status is set to 2 and we skip to the next codeword. 

\subsubsection*{Step 2 (Lines 4--11)}

For each found error location $(k,l) \in \mathcal{E}_{i,j}$, a
consistency check is performed. That is, one checks if the implied
component codeword $(k,l)$ corresponds to an anchor. If so,
$|\mathcal{L}_{k,l}|$ is the number of conflicts that this anchor is
already involved in. This number is then compared against a threshold
$\delta$. If $|\mathcal{L}_{k,l}| \geq \delta$, the anchor $(k,l)$ is
deemed unreliable and it is marked for backtracking by adding it to
the backtracking set $\mathcal{B}$. On the other hand, if
$|\mathcal{L}_{k,l}| < \delta$, the codeword $(i,j)$ is frozen by
changing its status to 3. Moreover, the conflict between the (now
frozen) codeword and the anchor is stored by modifying the respective
sets $\mathcal{L}_{i,j}$ and $\mathcal{L}_{k,l}$. Frozen codewords are
always skipped (in the loop of Algorithm 1) for the rest of the
decoding unless either the conflicting anchor is backtracked or any
bits in the frozen codeword change. 

%(see Algorithm \ref{alg:backtracking}) 
%(see Algorithm \ref{alg:error_correction})

\subsubsection*{Step 3 (Lines 12--15)} 

If the component codeword $(i,j)$ still has status 1, the bit flips
implied by $\mathcal{E}_{i,j}$ are consistent with all reliable
anchors, i.e., anchors that are involved in $\delta$ or fewer other
conflicts. If that is the case, the algorithm proceeds by applying the
error-correction step for codeword $(i,j)$, i.e., the bits $(i,j,k,l)$
corresponding to all error locations $(k,l) \in \mathcal{E}_{i,j}$ are
flipped. The error-correction step is implemented in Algorithm
\ref{alg:error_correction} and described in detail in Section
\ref{sec:error_correction}. Afterwards, the codeword $(i,j)$ becomes
an anchor by changing its status to 0. 

\subsubsection*{Step 4 (Lines 16--17)}

The last step consists of backtracking all anchor codewords in the set
$\mathcal{B}$ (if there are any).  Roughly speaking, backtracking
involves the reversal of all previously applied bit flips of the
corresponding anchor. Moreover, the backtracked codeword loses its
anchor status.  The backtracking routine is implemented in Algorithm
\ref{alg:backtracking} and described in more detail in
Sec.~\ref{sec:backtracking} below. 

\subsection{Examples}

We now illustrate the above steps with the help of two
examples. For both examples, a component code with
error-correcting capability $t=2$ is assumed. Moreover, the conflict
threshold is set to $\delta=1$.

\begin{figure*}[t]
	\centering \subfloat[]{\includegraphics{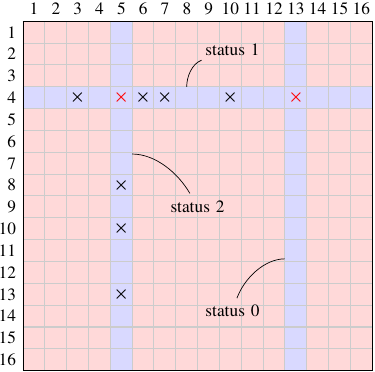}}\qquad
	\subfloat[]{\includegraphics{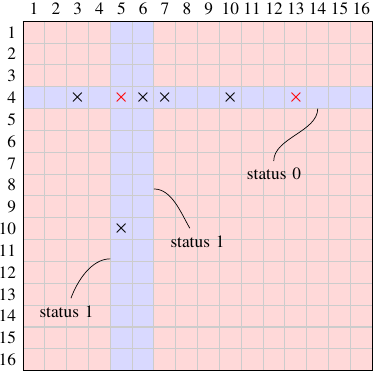}}\qquad
	\subfloat[]{\includegraphics{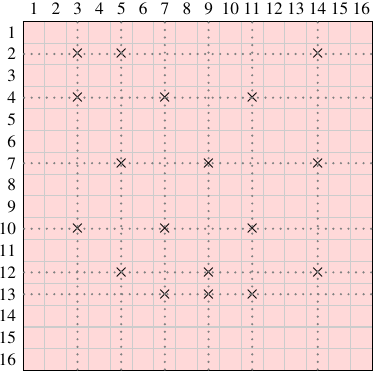}}\qquad
	\subfloat[]{\includegraphics{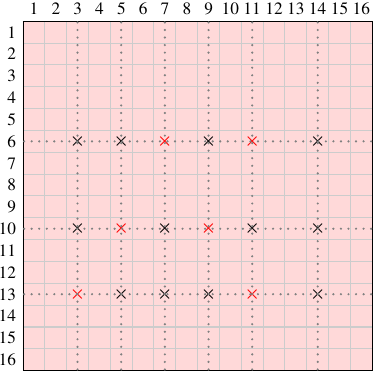}} \caption{(a)
	Error/status
	configuration illustrating how anchor codewords prevent
	miscorrections (see Example
	\ref{ex:miscorrection_avoidance}) , (b) Error/status configuration
	illustrating how backtracking is triggered for miscorrected anchors (see Example
	\ref{ex:backtracking}), (c) A minimal-size stopping set
	after algebraic-erasure post-processing (PP) assuming no miscorrections for $t=2$ and
	$e\in \{1, 2\}$, (d) A dominant error event for anchor-based
	decoding for $t=2$ and $e\in \{1,2\}$ in the error-floor regime.}
	\label{fig:code_arrays}
\end{figure*}

\begin{example}
	\label{ex:miscorrection_avoidance}
	 Consider the scenario depicted in Fig.~\ref{fig:code_arrays}(a).
	 Assume that we are at $(i,j) = (1,4)$, corresponding to a row
	 component codeword with status 1 and four attached errors shown by
	 the black crosses. The codeword is assumed to be miscorrected with
	 $\mathcal{E}_{3,4} = \{(2,5), (2,13)\}$ shown by the red crosses.
	 Codeword $(2,5)$ is assumed to have status 2 (i.e., \BDD failed in
	 the previous iteration with three attached errors) and therefore
	 the first consistency check is passed. However, assuming that the
	 codeword $(2,13)$ is an anchor without any other conflicts, i.e.,
	 $\mathcal{L}_{2,13} = \emptyset$, the codeword $(1,4)$ is frozen
	 during step 2. Hence, no bit flips are applied and the
	 miscorrection is prevented. The conflict is stored by updating the
	 two conflict sets as $\mathcal{L}_{1,4} = \{(2,3)\}$ and
	 $\mathcal{L}_{2,3} = \{(1,4)\}$, respectively. \demo
\end{example}

\begin{example}
	\label{ex:backtracking}
	Consider the scenario depicted in Fig.~\ref{fig:code_arrays}(b),
	where we assume that the codeword $(1,4)$ is a miscorrected anchor
	without conflicts (i.e., $\mathcal{L}_{1,4} = \emptyset$) and error
	locations $\mathcal{E}_{1,4} = \{(2,5), (2,13)\}$. Assume that we
	are at $(i,j) = (2,5)$. The codeword $(2,5)$ has status 1 and two
	attached error. Thus, \BDD is successful with
	 $\mathcal{E}_{2,1} = \{(1,4),(1,10)\}$.
	During step 2, the codeword $(2,5)$ is, however, frozen because there is a
	conflict with anchor $(1,4)$. After freezing the codeword, we have
	$\mathcal{L}_{1,4} = \{ (2,5)\}$ and $\mathcal{L}_{2,5} =
	\{(1,4)\}$. We skip to the next codeword $(2,6)$, which has status
	1. Again, \BDD is successful with
	$\mathcal{E}_{2,6} = \{(1, 4)\}$. The implied bit flip is
	inconsistent with the anchor $(1,4)$. However, since this anchor is
	already in conflict with codeword $(2,5)$ (and, hence,
	$|\mathcal{L}_{1,4}| = 1 = \delta$), the anchor is marked for
	backtracking and the error-correction step for bit $(2,6,1,4)$ will
	be applied. \demo
\end{example}

\subsection{Backtracking}
\label{sec:backtracking}

\IncMargin{1em}
%\restylealgo{boxed}
%\linesnumbered
\begin{algorithm}[t]
	\footnotesize
	\DontPrintSemicolon
	\SetKw{ShortFor}{for}
	\SetKw{Break}{break}
	\SetKw{MyWhile}{while}
	\SetKw{MyIf}{if}
	\SetKw{MySet}{set}
	\SetKw{MyElse}{else}
	\SetKw{MyCompute}{compute}
	\SetKw{KwEach}{each}
	\SetKw{KwAnd}{and}
	%\SetKwData{maxIter}{maxIter}
	\SetKwFunction{errProb}{error probability}
	%\EmptyLine
%	\KwIn{error graph $\mathcal{G}$} 
%	\KwOut{decoded error graph}
	%\BlankLine

	\For(\tcc*[f]{remove conflicts}){\KwEach $(k,l) \in \mathcal{L}_{i,j}$}{%
		remove $(k,l)$ from $\mathcal{L}_{i,j}$\;
		remove $(i,j)$ from $\mathcal{L}_{k,l}$\;
		\If(\tcc*[f]{no more conflicts}){\upshape $\mathcal{L}_{k,l}$ is empty}{%
			$(k,l).\algtext{status} \leftarrow 1$ \rightcomment{unfreeze the codeword}
		}%
	}%
	\For{\KwEach $(k,l) \in \mathcal{E}_{i,j}$}{%
		error-correction step for $(i,j,k,l)$ \rightcomment{see
		Alg.~\ref{alg:error_correction}}
	}%
	$(i,j).\algtext{status} \leftarrow 3$ \rightcomment{freeze codeword} 
	\caption{\small Backtracking anchor codeword $(i,j)$}
	\label{alg:backtracking}
\end{algorithm}
\DecMargin{1em}
%\vspace{-0.1cm}

In Example \ref{ex:backtracking}, we have encountered a scenario that
leads to the backtracking of a miscorrected anchor. The actual
backtracking routine is implemented in Algorithm
\ref{alg:backtracking}. First, all conflicts caused by the anchor are
removed by modifying the respective conflict sets. Note that all
codewords $(k,l) \in \mathcal{L}_{i,j}$ for anchor $(i,j)$ necessarily
have status 3, i.e., they are frozen. After removing conflicts, such
codewords may be conflict-free, in which case their status is changed
to $1$. After this, all previously applied bit flips are reversed. In
order to perform this operation, it is necessary to store the set
$\mathcal{E}_{i,j}$ for each anchor.  Finally, the codeword loses its
anchor status. In principle, the new codeword status can be chosen to
be either $1$ or $3$. However, backtracked anchors are likely to have
miscorrected. We therefore prefer to freeze the codeword by setting
its status to $3$ after the backtracking. 

\begin{remark}
	Since we do not know if an anchor is miscorrected or not, it is also
	possible that we mistakenly backtrack ``good'' anchors. Fortunately,
	this is unlikely to happen for long component codes because the
	additional errors due to miscorrections are approximately randomly
	distributed within the codeword \cite{Justesen2011}. This implies
	that additional errors of two (or more) miscorrected codewords
	rarely overlap. 
\end{remark}

\subsection{Error-correction Step}
\label{sec:error_correction}

\IncMargin{1em}
%\restylealgo{boxed}
%\linesnumbered
\begin{algorithm}[t]
	\footnotesize
	\DontPrintSemicolon
	\SetKw{ShortFor}{for}
	\SetKw{Break}{break}
	\SetKw{MyWhile}{while}
	\SetKw{MyIf}{if}
	\SetKw{MySet}{set}
	\SetKw{MyElse}{else}
	\SetKw{MyCompute}{compute}
	\SetKw{KwEach}{each}
	\SetKw{KwAnd}{and}
	\SetKw{KwNot}{not}
	%\SetKwData{maxIter}{maxIter}
	\SetKwFunction{errProb}{error probability}
	%\EmptyLine
%	\KwIn{error graph $\mathcal{G}$} 
%	\KwOut{decoded error graph}
	%\BlankLine

	%\If(\tcc*[f]{codeword is decodable}){$(i,j).\algtext{status} = 1$}{%
		%$\mathcal{B}.\algtext{clear}()$ \rightcomment{reset backtracking list}%
	\If{\KwNot $((i,j).\algtext{status} \,\,= 0$ \KwAnd
	$(k,l).\algtext{status} \,\,= 0)$  }{%
		flip the bit $(i,j,k,l)$\; %\rightcomment{freeze codeword}%
		%$R \leftarrow (k,l).\algtext{decode}$ \rightcomment{R indicates
		%success (1) or failure (0)}%
		\uIf{$(k,l).\algtext{status} = 2$}{%
			$(k,l).\algtext{status} \leftarrow 1$%
		}
		\ElseIf{$(k,l).\algtext{status} = 3$}{%
			$(k,l).\algtext{status} \leftarrow 1$\;%
			\For(\tcc*[f]{remove conflicts}){\KwEach $(k',l') \in \mathcal{L}_{k,l}$}{%
				remove $(k,l)$ from $\mathcal{L}_{k',l'}$\;
				remove $(k',l')$ from $\mathcal{L}_{k,l}$\;
			}%
		}%
	}%
	%
	% old code:
	%
%	\If{$(k,l).\algtext{status} \,\,!\!= 0$}{%
%		flip the bit $(i,j,k,l)$\; %\rightcomment{freeze codeword}%
%		$R \leftarrow (k,l).\algtext{decode}$ \rightcomment{R indicates
%		success (1) or failure (0)}%
%		\uIf{$(k,l).\algtext{status} = 1$ \KwAnd $R = 0$}{%
%			$(k,l).\algtext{status} \leftarrow 2$%
%		}\uElseIf{$(k,l).\algtext{status} = 2$ \KwAnd $R = 1$}{%
%			$(k,l).\algtext{status} \leftarrow 1$%
%		}%
%		\ElseIf{$(k,l).\algtext{status} = 3$}{%
%			\eIf{$R = 0$}{%
%				$(k,l).\algtext{status} \leftarrow 2$\;%
%			}{%
%				$(k,l).\algtext{status} \leftarrow 1$\;%
%			}%
%			\For(\tcc*[f]{remove conflicts}){\KwEach $(k',l') \in \mathcal{L}_{k,l}$}{%
%				remove $(k,l)$ from $\mathcal{L}_{k',l'}$\;
%				remove $(k',l')$ from $\mathcal{L}_{k,l}$\;
%			}%
%		}%
%	}%
	\caption{\small Error-correction step for bit $(i,j,k,l)$}
	\label{alg:error_correction}
\end{algorithm}
\DecMargin{1em}
%\vspace{-0.1cm}

The error-correction step is implemented in Algorithm
\ref{alg:error_correction}. The input is a parameter tuple $(i,j,k,l)$
where $(i,j)$ is the codeword that initiated the bit flip and $(k,l)$
is the corresponding codeword affected by it.  Note that Algorithm
\ref{alg:error_correction} can be reached from both the main routine
(Algorithm \ref{alg:main_routine}, lines 13--14) and as part of the
backtracking process (Algorithm \ref{alg:backtracking}, lines 6--7).
If the algorithm is reached via backtracking, it is possible that the
affected codeword $(k,l)$ is now an anchor. In this case, we use the
convention to trust the anchor's decision about the bit $(i,j,k,l)$
and not apply any changes. In all other cases, apart from actually
flipping the bit $(i,j,k,l)$ (line 2), error-correction triggers a
status change (lines 3--9). If the bit flip affects a frozen codeword,
the codeword is unfrozen and we remove the conflicts that the codeword
is involved in. 

%\cite{Justesen2011}

%For the algorithm to work well, a sufficiently large fraction of
%codewords at each position should be ``good'' anchors.  
%However, when the decoding window shifts and a new block is added, no
%anchors exist at the last position $W-1$. We found that it is
%therefore beneficial to artificially restrict the error-correcting
%capability of these component codes in order to avoid anchoring too
%many miscorrected codewords. For example, for $t=2$, all component
%codes at position $W-1$ are treated as single-error-correcting. This
%restriction reduces the probability of miscorrecting a component code
%by roughly a factor of $n$, which is significant for long component
%codes\cite{Justesen2011}. Note that due to the window decoding, we are
%merely gradually increasing the error-correction capability: once the
%decoding window shifts, the component codes shift as well and they are
%then decoded with their full error-correcting capability. 

\subsection{Generalized Product Codes}
\label{sec:gpcs}

In principle, anchor-based decoding can be applied to an arbitrary
\GPC. Indeed, Algorithms 2--4 are independent of the underlying \GPC.
The global code structure manifests itself only through the set of
error locations $\mathcal{E}_{i,j}$. This set is defined in terms of
the affected component codewords, which implicitly uses the code
structure. 

Compared to \PCs, the main difference for other \GPCs is that
Algorithm \ref{alg:iterative_bdd} has to be replaced by a version that
is appropriate for the specific \GPC. For anchor-based decoding,
Algorithm \ref{alg:iterative_bdd} simply specifies the
order in which the component codewords are traversed during the
iterative decoding. 

\begin{example}
	We have considered anchor-based decoding of staircase codes in the
	conference version of this paper \cite{Haeger2017ecoc}. In that
	case, Algorithm \ref{alg:iterative_bdd} is replaced by a
	window decoding schedule, see
	\cite[Alg.~1]{Haeger2017ecoc}. \demo
%	Simulation results for staircase codes can also be found in
%	\cite{Haeger2017ecoc}.
\end{example}

Note that staircase codes have more than two types of component
codewords. In particular, the codeword types indicate the position of
the component codewords in the staircase code array.
For \GPCs that do not admit a description in terms of a finite number of
codeword types (e.g., tightly-braided block codes), one may simply use
the convention that each component codeword forms its own type. 

%The conventional decoding procedure for staircase codes uses a
%sliding window comprising $W$ received blocks $\mat{B}_k,
%\mat{B}_{k+1}, \dots, \mat{B}_{k+W-1}$. This is illustrated in
%Fig.~\ref{fig:staircase_array} for $W = 5$ and $a = 6$.  It is
%convenient to identify each component code in the window by a tuple
%$(i,j)$, where $i \in \{1, 2, \dots, W-1\}$ indicates the position
%relative to the current decoding window and $j \in \{1, 2, \dots,
%a\}$ enumerates all codes at a particular position. As an example,
%the component codes $(1,3)$ and $(4,4)$ are highlighted in blue in
%Fig.~\ref{fig:staircase_array}. 

%However, when the decoding window shifts and a new block is added, no
%anchors exist at the last position $W-1$. We found that it is
%therefore beneficial to artificially restrict the error-correcting
%capability of these component codes in order to avoid anchoring too
%many miscorrected codewords. For example, for $t=2$, all component
%codes at position $W-1$ are treated as single-error-correcting. This
%restriction reduces the probability of miscorrecting a component code
%by roughly a factor of $n$, which is significant for long component
%codes\cite{Justesen2011}. Note that due to the window decoding, we are
%merely gradually increasing the error-correction capability: once the
%decoding window shifts, the component codes shift as well and they are
%then decoded with their full error-correcting capability. 

Anchor-based decoding can also be applied to \GPCs that are based on
component codes with different lengths and/or error-correcting
capabilities. For example, \PCs can be defined such that different
component codes are used to protect the rows and columns of the code
array. More generally, the component codes may even vary across rows
and columns, leading to irregular PCs \cite{Hirasawa1984,Alipour2012}.
While Algorithms 2--4 are agnostic to such changes, it may be
beneficial to adopt different conflict thresholds $\delta$ for
different component codes in these cases.

\section{Simulation Results}
\label{sec:results}

In this section, we present and discuss simulation results  assuming
different \BCH component codes. For the conflict threshold, we tested
different values $\delta \in \{0,1,2,3\}$ for a wide variety of
component codes and \BSC crossover probabilities. In all cases,
$\delta = 1$ was found to give the best performance. Hence, $\delta =
1$ is assumed in the following. 

\subsection{BCH Codes with $t = 2$}

Double-error-correcting \BCH codes are of particular interest because
they can be decoded very efficiently in hardware \cite{Gorenstein1960,
Condo2017}. On the other hand, the performance of the resulting PCs is
significantly affected by miscorrections. This was shown in Example
\ref{ex:performance_gap} in Sec.~\ref{sec:performance}. 

Recall that Example \ref{ex:performance_gap} uses a \BCH component
code with parameters $(7,2,1)$ and $\ell = 10$ decoding iterations.
For these parameters, the \BER of the anchor-based decoding is shown
in Fig.~\ref{fig:performance_gap} by the green line (circles). The
algorithm closely approaches the performance of the idealized
iterative \BDD in the waterfall regime. Moreover, virtually
miscorrection-free performance is achieved in the error-floor regime.

In order to quantify the performance gain with respect to iterative
\BDD, we use the \NCG. To that end, assume that a coding scheme with
code rate $R$ achieves a \BER of $p_\text{out}$ on a \BSC with
crossover probability $p$. The \NCG (in dB) is then defined as
\begin{align}
	\text{NCG} \define 10 \log_{10} \left( R \frac{(Q^{-1}\left(
	p_{\text{out}}
	\right))^2}{ (Q^{-1}\left( p \right))^2 } \right).
\end{align}
The \NCG assumes binary modulation over an additive Gaussian noise
channel and measures the difference in required $E_b/N_0$ between
uncoded transmission and coded transmission using the coding scheme
under consideration. As an example, it can be seen from
Fig.~\ref{fig:performance_gap} that the iterative \BDD and the
anchor-based decoding achieve a \BER of $10^{-8}$ at approximately $p=
1.31 \cdot 10^{-2}$ and $p = 1.69 \cdot 10^{-2}$, respectively. The
code rate in both cases is $R = 0.78$. Hence, the respective NCGs are
given by $6.96\,$dB and $7.37\,$dB. The proposed algorithm thus
achieves a \NCG improvement of approximately $\Delta \text{NCG} =
0.4\,$dB, as indicated by the arrow in Fig.~\ref{fig:performance_gap}.

%Test $\delta = 2$ for the anchors in the 

As suggest in \cite{Justesen2011}, correcting only one error in the
first iteration of iterative \BDD gives some moderate performance
improvements. This trick can also be used in combination with the
anchor-based decoding. In that case, a decoding failure (in line 2 of
Algorithm \ref{alg:main_routine}) also occurs when \BDD is successful
but $|\mathcal{E}_{i,j}| = 2$. Since the status of such component
codewords is set to 2, it is important to reset the status to 1 after
the first iteration in order to continue the decoding with the full
error-correction capability. For the PC in Example
\ref{ex:performance_gap} with $\ell = 10$ iterations, we found that
correcting only one error in the first iteration does not lead to
noticeable performance improvements for the anchor-based decoding.
Some small improvements can be obtained, however, by increasing the
total number of iterations to $\ell = 15$ and correcting only one
error in the first 5 iterations. This is shown by the green dotted
line in Fig.~\ref{fig:performance_gap}. It is important to stress that
without the gradual increase of $t$, the BER for $\ell = 10$ and $\ell
= 15$ is virtually the same. Thus, the improvement shown in
Fig.~\ref{fig:performance_gap} is indeed due to the artificial
restriction of the error-correcting capability. 

\begin{figure}[t]
	\begin{center}
		\includegraphics{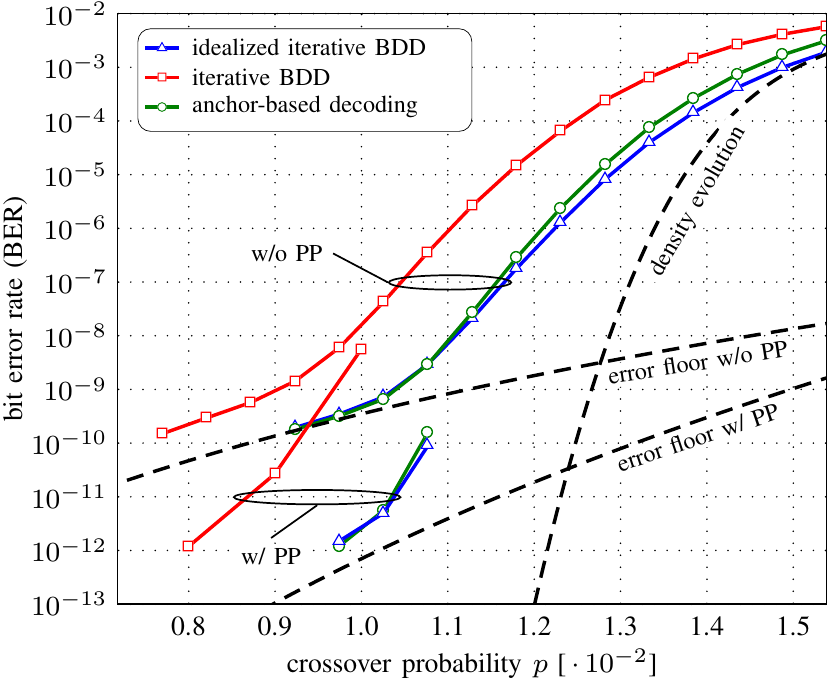}
	\end{center}
	\caption{Simulation results assuming a shortened BCH component code
	with parameters $(8,2,1)$. The resulting product code is considered
	in \cite{Condo2017}. Simulation data for iterative BDD including
	bit-flip-and-iterate post-processing (PP) was provided by the
	authors of \cite{Condo2017}. }
	\label{fig:pc_nu8_t2_s61_e1}
\end{figure}

Next, we provide a direct comparison with the results presented in
\cite{Condo2017}. In particular, the authors propose a hardware
architecture for a \PC that is based on a \BCH component code with
parameters $(8, 2, 1)$. The \BCH code is further shortened by 61 bits,
leading to an effective length of $n = 195$ and dimension $k=178$. The
shortening gives a desired code rate of $R = k^2/n^2 = 178^2/195^2
\approx 0.833$. The number of decoding iterations is set to $\ell =
4$. For these parameters, \BER results are shown in
Fig.~\ref{fig:pc_nu8_t2_s61_e1} for iterative \BDD, anchor-based
decoding, and idealized iterative \BDD (labeled ``w/o PP''). As
before, the outcome of \DE and the error floor prediction via
\eqref{eq:pc_error_floor} are shown by the dashed black lines as a
reference. Compared to the results shown in
Fig.~\ref{fig:performance_gap}, the anchor-based decoding approaches
the performance of the idealized iterative \BDD even closer and
virtually miscorrection-free performance is achieved for BERs below
$10^{-7}$. This can be attributed to the quite extensive code
shortening, which reduces the probability of miscorrecting compared to
an unshortened component code. 

\subsection{BCH Codes with $t > 2$}

For \BCH component codes with error-correcting capability larger than
2, the error floor is generally out of reach for our software
simulations. Hence, we focus on the performance improvements that can
be obtained in the waterfall regime. 

\begin{figure}[t]
	\begin{center}
		\includegraphics{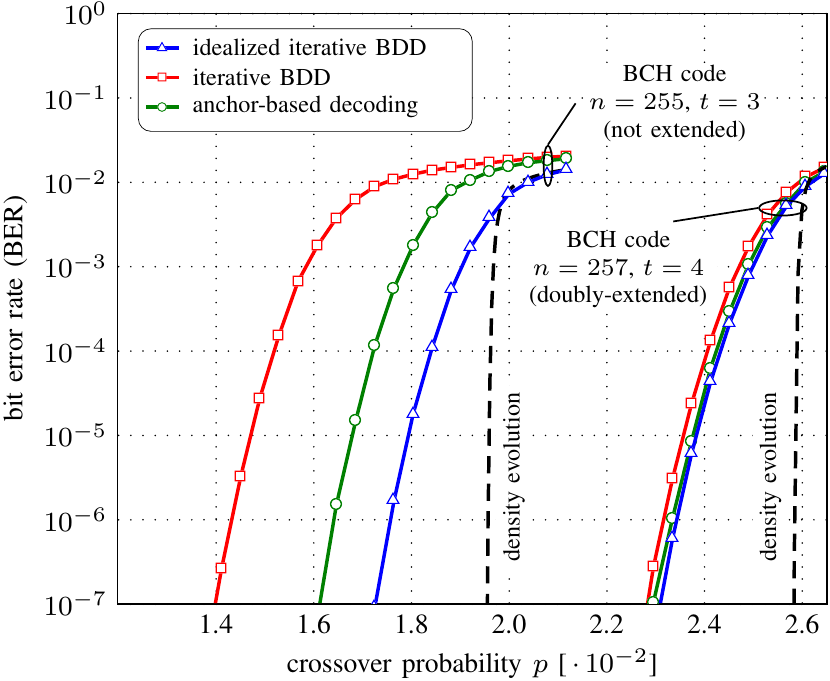}
	\end{center}
	\caption{Simulation results assuming two different BCH component codes with
	parameters $(8,3,0)$ and $(8,4,2)$. }
	\label{fig:pc_nu8_t3and4}
\end{figure}

In Fig.~\ref{fig:pc_nu8_t3and4}, we show the achieved BER for two
different PCs. The first PC is based on a BCH component code with
parameters $(8,3,0)$. For these parameters, miscorrections
significantly degrade the performance. Consequently, there is a large
performance gap between iterative \BDD and idealized iterative BDD.
The anchor-based decoding partially closes this gap and achieves a NCG
improvement of around $0.23\,$dB over iterative \BDD at $\text{BER} =
10^{-7}$. The second PC is based on a BCH component code with
parameters $(8,4,2)$. The miscorrection probability is reduced
approximately by a factor of 16 compared to the first PC. Hence, there
is only a small gap between iterative BDD and idealized iterative BDD.
The anchor-based decoding manages to close this gap almost completely.
The NCG improvement at $\text{BER} = 10^{-7}$ in this case is,
however, limited to around $0.01\,$dB. 

%As can be expected from \eqref{eq:miscorrection_probability}, the
%performance degradation with respect to idealized \BDD becomes less
%severe for larger values of $t$. 

%For BCH codes with larger error-correcting capabilities, qualitatively
%similar performance result are obtained. However, in that case, the 

%\subsection{Net Coding Gain Improvements}

%In particular, we consider BCH codes with $\nu = 9$, $t \in \{2, 3,
%4\}$ and $e \in \{0,1,2\}$, giving 9 different component codes in
%total. In Table~\ref{tab:overview}, we show the NCG improvements of
%anchor-based decoding over the conventional iterative \BDD. 
%
%The improvements range from $0.6\,$dB (for $t=2$ and $e=0$) to
%$0.02\,$dB (for $t=4$ and $e=2$). 

\section{Post-Processing}
\label{sec:post_processing}

If the anchor-based decoding terminates unsuccessfully, one may use
some form of \PP in order to continue the decoding. In this section,
we discuss two \PP techniques, which we refer to as
\emph{bit-flip-and-iterate} \PP and \emph{algebraic-erasure} \PP. Both
techniques have been studied before in the literature as a means to
lower the error-floor for various \GPCs assuming the conventional
iterative \BDD \cite{Sridharan2003, Jian2014, Emmadi2015, Condo2016b,
Mittelholzer2016, Holzbaur2017}. 

\subsection{Methods}

Let $\mathcal{F}_i$ denote the set of failed component codewords of
type $i$ after an unsuccessful decoding attempt. That is,
$\mathcal{F}_1$ and $\mathcal{F}_2$ are, respectively, row and column
codewords that still have nonzero syndrome after $\ell$ decoding
iterations. The intersection of these codewords defines a set of bit
positions according to
\begin{align}
	\label{eq:bad_intersection}
	\mathcal{I} = \{ (i,j,k,l) \,|\, (i,j) \in \mathcal{F}_1
	\text{ and } (k,l) \in \mathcal{F}_2 \}.
\end{align}
For \emph{bit-flip-and-iterate} \PP, the bits in the intersection
\eqref{eq:bad_intersection} are first flipped, after which the
iterative decoding is resumed for one or more iterations. The
intuition behind this approach is that the equivalent channel for the
bits in the intersection \eqref{eq:bad_intersection} after the
iterative decoding is essentially a \BSC with very high crossover
probability $p > 0.5$. Thus, the bit flipping converts this channel
into a \BSC where $p \ll 0.5$ and iterative decoding may then resolve
the remaining errors. Bit-flip-and-iterate \PP has been applied to
\PCs \cite{Sridharan2003, Condo2016b}, half-product codes
\cite{Mittelholzer2016}, and staircase codes \cite{Holzbaur2017}.  

%In fact, if $|\mathcal{F}_1| = |\mathcal{F}_2| = t+1$, the bit
%flipping is sufficient and no additional decoding iterations are
%necessary. 

For \emph{algebraic-erasure} \PP, the bits in the intersection
\eqref{eq:bad_intersection} are instead treated as erasures. An
algebraic erasure decoder is then used to recover the bits. Assuming
that there are no miscorrected codewords, algebraic-erasure \PP
provably succeeds as long as either $|\mathcal{F}_1| < \dmin$ or
$|\mathcal{F}_2| < \dmin$ holds. This type of \PP has been applied to
braided codes in \cite{Jian2014} and to half-product codes in
\cite{Emmadi2015}. 

\subsection{Post-Processing for Anchor-Based Decoding}
\label{sec:modifications}

In principle, the above \PP techniques can be applied after the
anchor-based decoding without any changes. However, it is possible to
improve the effectiveness of algebraic-erasure \PP by exploiting some
additional information that is available for the anchor-based
decoding. In particular, recall that it is necessary to keep track of
the error locations of anchors in case they are backtracked. If the
anchor-based decoding fails, these error locations can be used for the
purpose of \PP as follows. Assume that we have determined the sets
$\mathcal{F}_1$ and $\mathcal{F}_2$. Then, one can check for anchors
that satisfy the condition
\begin{align}
	\label{eq:suspicious_anchor_condition}
	\mathcal{E}_{i,j} \subset \mathcal{F} \quad \text{ and } \quad
	|\mathcal{E}_{i,j}| = t,
\end{align}
where $\mathcal{F} = \mathcal{F}_1 \cup \mathcal{F}_2$, i.e., anchors
that have corrected $t$ errors, where the error locations overlap
entirely with the set of failed component codewords. For the
algebraic-erasure PP, we found that it is beneficial to include such
anchors into the respective sets $\mathcal{F}_1$ and $\mathcal{F}_2$
(even though they have zero syndrome). This is because such component
codewords are likely to be miscorrected. 

Note that additional component codewords should only be included into
the sets $\mathcal{F}_1$ and $\mathcal{F}_2$ as long as
$|\mathcal{F}_1| < \dmin$ or $|\mathcal{F}_2| < \dmin$ remain
satisfied. Therefore, if there are more component codewords satisfying
\eqref{eq:suspicious_anchor_condition} than allowed by these
constraints, we perform the inclusion on a random basis. A better, but
also more complex, approach would be to perform the algebraic erasure
decoding multiple times with all possible combinations of component
codewords in the sets $\mathcal{F}_1$, $\mathcal{F}_2$ and those
satisfying \eqref{eq:suspicious_anchor_condition}. 

\begin{remark}
	For bit-flip-and-iterate PP, we found that the same strategy may, in
	fact, degrade the performance, in particular if the number of
	decoding iterations is relatively low. For small $\ell$, the
	decoding often terminates unsuccessfully with only a few remaining
	errors left. These errors are then easily corrected using the
	conventional bit-flip-and-iterate PP. In that case, it is
	counterproductive to include additional component codewords into the
	sets $\mathcal{F}_1$ and $\mathcal{F}_2$. 
%	In this case, the decoding may terminate with a few remaining errors
%	which would have been resolved, had we done one more iteration. In
%	such cases, it can be counterproductive to include additional
%	component codewords into 
\end{remark}

\subsection{Example for BCH Codes with $t = 2$}

Consider again the \PC based on a shortened \BCH code with parameters
$(8, 2, 1)$ studied in \cite{Condo2017} (see
Fig.~\ref{fig:pc_nu8_t2_s61_e1}).  The authors in \cite{Condo2017}
also consider the application of bit-flip-and-iterate PP in order to
reduce the error floor. The simulation data was provided to us by the
authors and the results are reproduced\footnote{The same results are
shown in \cite[Fig.~3]{Condo2017}.} for convenience in
Fig.~\ref{fig:pc_nu8_t2_s61_e1} by the red line (squares) labeled ``w/
PP''. 
For the anchor-based decoder, we propose to use instead the
algebraic-erasure \PP as described in the previous subsection. In
order to estimate its performance, we first consider algebraic-erasure
\PP for the idealized iterative \BDD without miscorrections. In this
case, the dominant stopping set is of size $18$ and involves 6 row and
6 column codewords.  An example of this stopping set is shown by the
black crosses in Fig.~\ref{fig:code_arrays}(c), where the involved
rows and columns are indicated by the dotted lines.  It is pointed out
in \cite{Holzbaur2017} that the multiplicity of such a stopping set
can be obtained by using existing counting formulas for the number of
binary matrices with given row and weight weight \cite{Wang1998}.  In
particular, there exist 297,200 binary matrices of size $6\times 6$
with uniform row and column weight 3 \cite[Table 1]{Wang1998}. This
gives a multiplicity of $M = 297,200 \binom{n}{6}^2$ for this stopping
set. The error floor can then be estimated using
\eqref{eq:pc_error_floor} with $\smin = 18$. This is shown in
Fig.~\ref{fig:pc_nu8_t2_s61_e1} by the dashed black line labeled
``error floor w/ PP'' and can be verified using the idealized
iterative \BDD including \PP. The performance of anchor-based decoding
including algebraic-erasure \PP is also shown in
Fig.~\ref{fig:pc_nu8_t2_s61_e1} and virtually overlaps with the
performance of idealized iterative \BDD for BERs below $10^{-11}$.
Overall, the improvements translate into an additional \NCG of around
$0.2\,$dB at a \BER of $10^{-12}$ over iterative \BDD
with bit-flip-and-iterate PP.

Without the modification described in Sec.~\ref{sec:modifications},
the performance of algebraic-erasure \PP assuming the anchor-based
decoding would be slightly decreased.  Indeed, we found that a
dominant error event of the anchor-based decoder for $t=2$ and $e \in
\{1, 2\}$ is such that 3 row (or column) decoders miscorrect towards
the same estimated error pattern $\hat{\vect{n}}$ of weight 6. This
scenario is illustrated in Fig.~\ref{fig:code_arrays}(d). We did not
encounter similar error events for the conventional iterative \BDD.
This indicates that the anchor-based decoder introduces a slight bias
towards an estimated error pattern, once it is ``anchored''. For the
error event shown in Fig.~\ref{fig:code_arrays}(d), 6 column codewords
are not decodable, whereas all row codewords are decoded successfully
with a zero syndrome. This implies that the set $\mathcal{F}_2$ is
empty and therefore the bit intersection \eqref{eq:bad_intersection}
is empty as well. Hence, conventional \PP (both bit-flip-and-iterate
and algebraic-erasure) would fail. On the other hand, with high
probability condition \eqref{eq:suspicious_anchor_condition} holds for
all 3 miscorrected row codewords. Condition
\eqref{eq:suspicious_anchor_condition} may also hold for correctly
decoded anchors. However, there is no harm in including up to two
additional correctly decoded row codewords into the set
$\mathcal{F}_2$.

\section{Implementation Complexity}
\label{sec:complexity}

One of the main advantages of iterative \BDD compared to
message-passing decoding is the significantly reduced decoder data
flow \cite{Smith2012a}. In this section, we briefly review
the product decoder architecture in \cite{Smith2012a} and discuss the
potential increase in implementation complexity for the proposed
anchor-based decoding. However, the design of a full hardware
implementation is beyond the scope of this paper. 

%With this in mind, our main goal is show that anchor-based decoding
%has essentially the same data flow requirements as iterative BDD. On
%the other hand, additional storage and a more complicated control
%unit are needed. 

The product decoder architecture discussed in \cite{Smith2012a}
consists of three main parts or units (cf.~\cite[Fig.~3]{Smith2012a}):
a data storage unit for the product code array, a syndrome storage
unit, and a BCH component decoder unit. Based on this architecture,
the authors argue that the internal data flow (in bits/s) between
these units can be used as a surrogate for the implementation
complexity of iterative BDD of PCs. The main contributors to the
overall decoder data flow are as follows:

\begin{itemize}
	\item Initially, the syndromes for all component codewords have to
		be computed and stored in the syndrome storage based on the
		received data bits.

	\item During iterative decoding, syndromes are loaded from the
		syndrome storage and used by BCH component decoder unit for the
		component decoding. %Per product codeword: $2 r n v$

	\item After a successful component decoding, the syndromes in the
		syndrome storage are updated based on the found error locations.
		%$2 t n r v$
		Moreover, the corresponding bits in the data storage unit are
		flipped. % $2 t n v 2\log2(n)$

\end{itemize}

A key aspect of this architecture is that information between
component codewords is exchanged entirely through their syndromes.
This is very efficient at high code rates: in this case, syndromes can
be seen as a compressed representation of the component codeword.
Moreover, each successful component decoding affects at most $t$
syndromes. 

%It is also shown \cite{Smith2012a} that the data flow caused by the
%actual \BDD for BCH component codes with $t=3$ can be neglected. 

In principle, anchor-based decoding can also operate entirely in the
syndrome domain, thereby leveraging the same syndrome compression
effect as iterative BDD. In particular, the initial syndrome
computation phase can be performed in the same fashion as for
iterative BDD. The syndrome loading occurs before executing line 2 of
Algorithm \ref{alg:main_routine} and syndrome updates are triggered
due to line 2 of Algorithm \ref{alg:error_correction}. 

While the syndrome domain implementation can be kept intact, in the
following we comment on some of the implementation differences between
iterative BDD and anchor-based decoding. 

%Our considerations are mainly based on the specific hardware
%architectures that are discussed in \cite{Smith2012a} and
%\cite{Condo2017}. 

\subsubsection{Component code status}

Anchor-based decoding uses a status value for each component code.
Status changes occur after BDD (lines 15 and 19 in Algorithm
\ref{alg:main_routine}), after backtracking (line 8 in Algorithm
\ref{alg:backtracking}), and after applying bit flips (lines 3--6 in
Algorithm \ref{alg:error_correction}). This leads to an apparent
complexity increase compared to a straightforward implementation of
iterative BDD. On the other hand, even for iterative \BDD, it is
common to introduce some form of status information for each component
codeword. For example, a status flag is often used to indicate if the
syndrome for a particular component code changed since the last
decoding attempt \cite{Smith2012a}. This is done in order to avoid
decoding the same syndrome multiple times. The product decoder
architecture in \cite{Condo2017} also features a status flag to
indicate the failure of a particular component code in the last
decoding iteration. Therefore, the slightly more involved status
handling for the anchor-based decoding should not lead a drastic
complexity increase compared to practical implementations of iterative
BDD. 

%What is not discussed in \cite{Smith2012a} is the presence of a
%control unit that handles the scheduling in the iterative decoding.

\subsubsection{Error locations}

Additional storage is needed to keep track of the error locations
$\mathcal{E}_{i,j}$ for each anchor codeword $(i,j)$ in case of
backtracking (lines 6--7 in Algorithm \ref{alg:backtracking}). Since
each individual error location can be specified using
$\ceil{\log_2(n)}$+1 bits, the total extra storage for all error
locations required is $2 n t (\ceil{\log_2(n)}$+1) bits. 

\subsubsection{Conflict sets}

Additional storage is also needed to store the conflicts between
component codewords. For a $t$-error-correcting component code, there
can be at most $t$ conflicts for each frozen component codewords.
Moreover, for a conflict threshold of $\delta = 1$, it is sufficient
to keep track of a single conflict per anchor codeword. Taking the
larger of these two values, the conflict set size therefore has to be
$t$. The extra storage required is thus the same as for the error
locations, i.e., $2 n t (\ceil{\log_2(n)}+1)$ bits. One possibility to
reduce the required storage is to only keep track of a single conflict
for each frozen component codeword and ignore other conflicts. This
would also lead to a very simple implementation of the loops in
Algorithm \ref{alg:backtracking} (line 1) and Algorithm
\ref{alg:error_correction} (line 7). On the other hand, this may also
lead to a small performance loss. 

%\subsubsection{Consistency checks}

%\subsubsection{Backtracking}
%
%In its current form, backtracking of an anchor codeword occurs
%immediately after the anchor is deemed unreliable due to too many
%conflicts (lines 16-17 in Algorithm \ref{alg:main_routine}). This may
%pose a potential issue for architectures that rely on pipelining the
%BCH component decoder stages \cite{Condo2017}. In this case, an
%alternative approach is to separate the status change 
%
%One
%potential issue with this is that backtracking does not occur
%frequently
%
%An alternative would be as follows. Rather than backtracking
%immediately, one could set the status to 1 and introduce an additional
%backtracking flag. In this case, the bit flipping would occur 
%
%the codeword would not be decoded in
%the next iteration but only the bit flipping would occur in the next
%iteration. 
%
%Note that the anchor . We implemented this strategy in software. For
%the PC discussed in Example \ref{ex: }, we found no noticeable
%performance difference compared to . 
%
%This indicates that there is some implementation flexibility regarding
%the handling of backtracking in a heavily pipelined and parallelized
%hardware architecture. 

%\subsection{Backtracking}
%
%One potential issue 

%For example, the proposed decoder architecture in \cite{Condo2017}
%uses an array of $13$ parallel BCH decoder units for a \PC with
%component code of length $n = 195$. 

%Note that in Algorithm 1, the \BDD of component codewords of the same
%type is serialized. That is, rather than applying \BDD to all row or
%column codewords at once, the decoding is done sequentially. 

\section{Conclusion}
\label{sec:conclusion}

We have shown that the performance of product codes can be
significantly improved by adopting a novel iterative decoding
algorithm. The proposed algorithm uses anchor codewords to reliably
detect and prevent miscorrections. Depending on the component code
parameters and the BSC crossover probability, anchor-based decoding
can closely approach the performance of idealized iterative BDD where
a genie avoids all miscorrections. Moreover, the algorithm can be
applied to a large variety of related code classes that are used in
practice, e.g., staircase or braided codes. 

%\bibliographystyle{IEEEtran}
%\bibliography{$HOME/Dropbox/lib/bibtex/library_mendeley}

%\appendix
\appendices

\section{Density Evolution for Product Codes}
\label{app:de}

In this appendix, we briefly describe how to reproduce the \DE results
that are shown in Figs.~\ref{fig:performance_gap},
\ref{fig:pc_nu8_t2_s61_e1}, and \ref{fig:pc_nu8_t3and4}. 

Consider a BCH component code with parameters $(\nu, t, e)$. The goal
is to predict the waterfall \BER of the \PC $\mathcal{P}(\mathcal{C})$
under iterative \BDD as a function of the \BSC crossover probability
$p$. Recall that the component code length is $n = 2^\nu-1+e$ and the
number of iterations is $\ell$, where each iteration consists of two
half-iterations with row and column codewords being decoded
separately. We define $c = pn$, $L = 2$, and $\etab
=\left(\begin{smallmatrix} 0 & 1\\ 1 & 0
\end{smallmatrix}\right)$. Further, let $\mathcal{A}^{(l)} = \{1\}$ for 
$l$ odd and $\mathcal{A}^{(l)} = \{2\}$ for $l$ even, where $l \in
[2\ell]$. Finally, let $\Psi_{\geq t}(\lambda) = 1 - e^{-\lambda}
\sum_{i=0}^{t-1} \frac{\lambda^i}{i!}$ be the tail probability of a
Poisson random variable with mean $\lambda$. With these definition, we
recursively compute
\begin{align}
	x_i^{(l)} = \begin{cases} 
		\Psi_{\geq t}
		\left( 
			\frac{c}{L} \sum_{j=1}^L \eta_{i,j} x_{j}^{(l-1)} 	
			\right) &\quad \text{if }i \in \mathcal{A}^{(l)}\\
			x_i^{(l-1)} &\quad \text{otherwise}
	\end{cases}
\end{align}
for $i \in [L]$ and $l = 1,2, \dots, 2\ell$, using $x_i^{(0)} = 1$, $i
\in [L]$, as initial values. Collecting all final values in a
vector as $\vect{x} = ({x}_1^{(2\ell)}, \dots, {x}_L^{(2\ell)})$,
the \BER is then approximated as
\begin{align}
	\text{BER}(p) \approx p \frac{\vect{x} \mat{\eta} \vect{x}^\transpose}{
	\|\mat{\eta}\|_\text{F}^2
	},
\end{align}
where $\|\mat{\eta}\|_\text{F}^2$ is the number of 1s in $\mat{\eta}$. 

%Technically, the above procedure implicitly assumes that \BDD is
%applied simultaneously to \emph{all} component codes in parallel. On
%the other hand, for the decoding schedule that alternates between row
%and column component codes, one may introduce a schedule as follows. 
%Let $\mathcal{A}^{(l)}$ for $l \in [\ell]$. 

The above procedure can be applied to compute the asymptotic
performance for a wide variety of \GPCs by simply adjusting the code
parameters $L$ and $\mat{\etab}$, and the decoding schedule
$\mathcal{A}^{(l)}$. For example, for staircase codes, the matrix
$\mat{\eta}$ is an $L \times L$ matrix with entries $\eta_{i,i+1} =
\eta_{i+1,i} = 1$ for $i \in [L-1]$ and zeros elsewhere. For more
details on this topic, we refer the interested reader to
\cite{Haeger2017tit}.

\end{document}